\UseRawInputEncoding
\documentclass[prx,aps,twocolumn,superscriptaddress]{revtex4}

\usepackage{dcolumn}
\usepackage{bm}
\usepackage{amssymb}
\usepackage{color}
\usepackage[pdftex]{graphicx}

\usepackage{amsmath}
\usepackage[linesnumbered,ruled]{algorithm2e}

\usepackage{url}
\usepackage[section]{placeins}
\usepackage[colorlinks=true,linkcolor=blue,citecolor=blue]{hyperref}

\begin{document}


\title{Kinetics of thermal Mott transitions in the Hubbard model}

\author{Gia-Wei Chern}
\affiliation{Department of Physics, University of Virginia, Charlottesville, VA 22904, USA}

\date{\today}

\begin{abstract}
We present the first-ever multi-scale dynamical simulation of the temperature-controlled Mott metal-insulator transition in the Hubbard model. By integrating advanced electronic structure method and an efficient Gutzwiller/slave-boson solver into molecular dynamics simulations, we demonstrate that the transformation from a correlated metal to the Mott insulating phase proceeds via the nucleation and growth of the Mott droplets. Moreover, the time evolution of the Mott volume fraction is found to follow a universal transformation kinetics. We show that after an initial incubation period, the early stage of the phase transformation is characterized by a constant nucleation rate and an interface-controlled cluster growth mechanism, consistent with the classical theory developed by Kolmogorov, Johnson, Mehl, and Avrami. This is followed by a novel intermediate stage of accelerated phase transformation that is significantly different from the prediction of the classical theory. Moreover, the cluster-growth dynamics in this intermediate stage exhibits an unexpected avalanche behavior, similar to the Barkhausen noise in magnetization dynamics, even in the absence of quenched disorder. Detailed structural characterization further uncovers a universal correlation function for the transient mixed-phase states of the Mott transition. We also discuss implications of our findings for spatially resolved measurements of Mott metal-insulator transition obtained in recent nano-imaging experiments. 
\end{abstract}

\maketitle

\section{Introduction}
\label{sec:intro}

Mott metal-insulator transition (MIT) continues to be an important subject in modern condensed matter physics even after more than sixty years of study~\cite{mott90,imada98,dobrosavljevic12}. In particular, the unusual properties of the correlated metallic state near the MIT, sometimes also called the bad metal~\cite{emery95}, have attracted considerable attention both experimentally and theoretically~\cite{limelette03,bartosch10,itou17,pustogow18,terletska11,sordi12,vucicevic15,jawad15,lenz16,imada05}. Conventionally, the effort has been focused on the thermodynamic, optical, magnetic, and transport properties in the vicinity of the Mott transition. These are usually achieved experimentally through bulk or macroscopic measurements. Coupled with the advances in many-body techniques such as the quantum Monte Carlo simulations~\cite{leblanc15,motta17,kent18} and the dynamical mean-field theory (DMFT)~\cite{georges96,kotliar06,gull11}, significant progress has been made toward understanding the microscopic mechanisms and the quantum criticality of the~MIT.

An important aspect that has received considerable attention recently is the {\em complexity} of MITs in real materials~\cite{dagotto05}. 
For example, several near-field spectroscopy and nano-imaging experiments~\cite{bonnell12,atkin12,liu17} have revealed highly inhomogeneous electronic states with complex nano-scale textures during the first-order metal-insulator transformation in correlated electron systems~\cite{qazilbash07,liu13,madan15,lupi10,qzailbash11,mcleod16,stinson18,mattoni16,ronchi18,preziosi18,singer18}. In fact, MIT in most functional materials, such as vanadium dioxide, is {\em discontinuous} at ambient conditions.
Understanding the structural and dynamical properties of the mesoscopic patterns during the first-order MIT is not only of fundamental interest, but also has important technological implications.  

In the past decade, considerable experimental effort has been devoted to understanding the properties of the heterogeneous electronic states in the vicinity of the Mott transition. For instance, the nano-scale phase separation in several temperature-driven MITs was shown to be consistent with the nucleation and percolation scenario~\cite{qazilbash07,lupi10,mcleod16,liu13}, similar to that observed in the colossal magnetoresistant materials (CMR)~\cite{uehara99,fath99,zhang02}. On the theoretical side, however, a {\em microscopic} modeling of these emergent dynamical textures is still lacking. The difficulty is partly due to the multiscale and multi-faceted nature of the MIT kinetics. On one hand, large-scale dynamical simulation is required in order to describe the complex structures of the mixed-phase states in the Mott transition. On the other hand, accurate modeling of the electron correlation effects requires advanced many-body techniques~\cite{leblanc15,motta17,kent18,georges96,kotliar06,gull11}, most of which are computationally too expensive to be directly combined with large-scale simulations.  One exception is the Gutzwiller approach~\cite{gutzwiller63,gutzwiller64,gutzwiller65}, which allows for computational efficiency similar to the Hartree-Fock methods or the density functional theory (DFT). Importantly, the Gutzwiller method captures essential correlation effects such as the bandwidth renormalization and electron localization~\cite{brinkman70}, and offers perhaps the only feasible way for large-scale real-space simulations of MIT.

The mesoscale electronic inhomogeneities associated with the MIT are reminiscent of the complex patterns observed in numerous first-order phase transitions of, e.g. binary alloys, polymer mixtures, and many other physical systems~\cite{langer80,gunton83,puri09}. The study of the first-order transition kinetics has a long history~\cite{binder87,onuki02}. Microscopically, the kinetic Ising model with, e.g. Glauber-type dynamics, has been extensively used to simulate the phase transformation  dynamics~\cite{marro79,phani80,sadiq83,stauffer82,shneidman99,shneidman99b,wonczak00,soisson00,brendel05,ryu10,binder16}. The Ising model or the liquid-gas transition are also often invoked to model the Mott MIT by, e.g. associating the gas and liquid phases with the metallic phase and Mott insulator, respectively~\cite{limelette03,bartosch10,itou17,liu16,bar18,papanikolaou08}. Several works have indicated that Mott transitions likely belong to the Ising universality class. One objective of this work is to investigate whether the kinetic Ising model can also account for the nucleation dynamics of the Mott transition.

The phase transformation dynamics is conventionally formulated in terms of the appropriate order parameters that characterize the phase transition~\cite{binder87,onuki02}. However, since  the MIT is a transition between different transport behaviors and is not characterized by broken symmetries, there is no clear choice of the order parameter. For Mott transitions, however, making use of the analogy with the liquid-gas transition, one can introduce an order parameter based on, e.g. the density of doublons, or doubly occupied sites~\cite{lee17}. Since the Gutzwiller method can be viewed as a mean-field theory of the Mott transition within the Kotliar-Ruckenstein slave-boson framework~\cite{kotliar86}, it offers a systematic way to construct the relevant order parameters for a Mott transition. Specifically, these order parameters correspond to the coherent state amplitude~$\Phi$ of the slave bosons~\cite{lanata12,lanata15}. For example, the slave-boson variables in a single-band Hubbard model are represented by $\Phi = (e, p_{ \uparrow}, p_{ \downarrow}, d)$; the square of its elements gives the probability of the empty, singly occupied (up and down spin), and doubly occupied states, respectively~\cite{kotliar86}. It thus encodes information about the local electron density $n_\alpha = |p_\alpha|^2$ as well as the double occupancy $D = |d|^2$. This formulation will be particularly useful for MIT in multi-orbital systems, where a large number of local electron states and potential order-parameters have to be considered~\cite{bunemann98,lanata15,lanata17}. 

Phenomenologically, the dynamics of the order-parameter field depends on the conservation laws as well as the relevance of other hydrodynamic variables~\cite{hohenberg77}. The simplest dynamics for a non-conserved order parameter $\Phi(\mathbf r, t)$,  as in the case of Mott MIT, is the time-dependent Ginzburg-Landau (TDGL) equation~\cite{binder73,valls90,iwamatsu08}
\begin{eqnarray}
	\label{eq:TDGL}
	\frac{\partial \Phi}{\partial t} = -\Gamma \frac{\partial \mathcal{E}}{\partial \Phi^\dagger} + \eta(\mathbf r, t).
\end{eqnarray}
Here $\mathcal{E}[\Phi]$ is an effective free-energy functional, $\Gamma$ denotes the inverse damping coefficient, and $\eta(\mathbf r, t)$ is a Gaussian white noise. In the case of the kinetic Ising model with Glauber dynamics, the TDGL corresponds to the coarse-grained approximation of the mean-field master equation~\cite{puri09}, and is often used in simulations of liquid-gas or magnetic transitions with non-conserved kinetics~\cite{binder87,onuki02}. The stochastic $\eta$ term, which plays an important role in the nucleation process, comes from the thermal fluctuations of the microscopic variables that are ``integrated'' out in the above formulation.  Although the TDGL has also been used in modeling the hysteresis dynamics of the MIT~\cite{bar18}, its microscopic foundation has yet to be carefully examined. Indeed, a formal derivation of the dynamical equation for the order parameter of the Mott transition requires integrating out the microscopic electron and lattice degrees of freedom. This is generally very difficult if not impossible, and phenomenological expressions are often used in the kinetics simulations. 

In this paper, we perform the first-ever microscopic simulation of the thermal Mott transition in the Hubbard model by keeping both lattice and electron degrees of freedom. We employ the Langevin molecular dynamics (MD) method to simulate the lattice fluctuations, which provide the relaxation mechanism and the source of stochastic force for the order parameter field. To explicitly include the microscopic correlation effects, the electron degrees of freedom are {\em dynamically} integrated out during the MD simulation. This is achieved by solving the Hubbard model using an efficient Gutzwiller method at every MD time-step. 
We~show that the temperature-driven Mott transition is described by the nucleation and growth scenario, consistent with several recent nano-imaging experiments. We further characterize the different stages of the transformation dynamics and the structure of the intermediate mixed-phase states.

The rest of the paper is organized as follows. Section~\ref{sec:GMD} presents the new quantum MD scheme based on the Gutzwiller method, and the Hubbard model on a deformable lattice to be studied in this work. Section~\ref{sec:thermal-mott} discusses the finite-temperature phase diagram of the Hubbard model and outlines the Gutzwiller MD simulations of the thermal Mott transition driven by temperature quenches. In Section~\ref{sec:nucleation}, the evolution of the cluster-size distribution from our MD simulations is discussed within the classical theory of nucleation-and-growth kinetics~\cite{becker35,zeldovich43,frenkel46}. Section~\ref{sec:interface} presents numerical evidences for an interface-controlled cluster-growth mechanism at the early stage of the phase transformation. Comparison of the transition kinetics with the classical Kolmogorov, Johnson, Mehl, and Avrami model~\cite{kolmogorov37,avrami39,avrami40,johnson39} is also discussed. In Section~\ref{sec:avalanche}, we discuss the novel intermediate stage of the transformation that is characterized by an avalanche cluster-growth behavior. Section~\ref{sec:structure} presents the numerical structure factor and two-point correlation function of the mixed-phase states during the Mott transition. And finally, Section~\ref{sec:outlook} concludes the paper with an outlook of future directions. 

\section{Gutzwiller molecular dynamics}
\label{sec:GMD}

We consider the half-filled Hubbard model on a compressible triangular lattice
\begin{eqnarray}
	\label{eq:H}
	\mathcal{H} = \sum_{\langle ij \rangle, \,\alpha } t(\mathbf u_i - \mathbf u_j) \left(c^\dagger_{i,\alpha} c^{\;}_{j, \alpha}  +  c^\dagger_{j,\alpha} c^{\;}_{i, \alpha} \right) 
	+ U \sum_i n_{i,\uparrow} n_{i,\downarrow} \nonumber \\
	+ \frac{K_0}{2} \sum_i |\mathbf u_i|^2 + \frac{K_1}{2} \sum_{\langle ij \rangle} \left[\hat{\mathbf e}_{ij} \cdot (\mathbf u_j - \mathbf u_i) \right]^2 + \sum_i \frac{|\mathbf p_i|^2}{2m}. \quad
\end{eqnarray}
Here $c^\dagger_{i, \alpha}$ is the creation operator of electron with spin $\alpha = \uparrow, \downarrow$ at site-$i$, $n_{i, \alpha} = c^\dagger_{i,\alpha} c^{\;}_{i, \alpha}$ is the electron number operator, $U$ is the Hubbard repulsion parameter, $K$ is a elastic constant, $\mathbf p_i$ is the momentum operator, $m$ is the mass of the atom, $\mathbf u_i$ denotes the displacement vector of the $i$-th site, i.e. $\mathbf r_i = \mathbf r^{(0)}_i + \mathbf u_i$, and $\hat{\mathbf e}_{ij}$ is a unit vector pointing from site-$i$ to $j$. We assume the following dependence of hopping integral on the displacements 
\begin{eqnarray}
	\label{eq:t_ij}
	t_{ij} = t(\mathbf u_i - \mathbf u_j) = t^{(0)}_{ij} \left[1 + g \, \hat{\mathbf e}_{ij} \cdot (\mathbf u_j - \mathbf u_i) \right],
\end{eqnarray}
where $t^{(0)}_{ij}$ is the bare hopping constant, and $g$ is the electron-phonon coupling. The Hamiltonian~(\ref{eq:H}) with $t_{ij}$ given by Eq.~(\ref{eq:t_ij}) is called the Peierls-Hubbard model~\cite{mazumdar83,hirsch83}, which has served as the basic platform for studying the interplay between the Peierls instability and electron correlation~\cite{tang88,fehske92,yuan02}. Here we choose the frustrated triangular lattice to avoid the unnecessary complexity due to Fermi-surface nesting and the associated spontaneous symmetry-breaking or structural transition. 
As discussed in Sec.~\ref{sec:intro}, the lattice degrees of freedom here serve to introduce the stochastic dynamics for the order parameter of the Mott transition.

The experimental time scale of temperature-driven Mott transition, which is also the time scale $\tau_\Phi$ for the order-parameter dynamics, is usually much longer than the characteristic time of lattice fluctuations. On the other hand, the relaxation time of electrons $\tau_e$, which ranges from a few tens to hundred femtoseconds, is short compared with the lattice time scale $\tau_L$ of the picosecond order. This separation of time scales allows us to apply the Born-Oppenheimer (BO) approximation in our MD simulations. Indeed, BO approximation is often employed in {\em ab inito} or quantum MD method~\cite{marx09,plasienka17}. The state-of-the-art DFT-based quantum MD, however, cannot describe the Mott transition. In order to capture the electron correlation effects, here we employ a new quantum MD scheme in which the atomic forces are computed on-the-fly from the Gutzwiller solution of Hubbard-type models~\cite{chern17}. Applying the Gutzwiller MD to a Hubbard liquid model, we have recently demonstrated, for the first time, the MD simulations of Mott MIT in an atomic liquid~\cite{chern17}.

The atomic displacements $\mathbf u_i$ in the finite-temperature MD simulations are governed by the Newton equation of motion with the Langevin thermostat~\cite{allen89,rapaport11}
\begin{eqnarray}
	\label{eq:langevin}
	m \ddot{\mathbf u}_i = \mathbf f^l_{i} + \mathbf f^e_{i} -\gamma \dot{\mathbf u}_i + \bm \xi_i(t),
\end{eqnarray} 
where $\gamma$ is the damping coeficient, $\bm\xi_i(t)$ are stochastic forces with statistical properties $\langle \xi^a_i \rangle = 0$ and $\langle \xi^a_i(t) \xi^b_j(t') \rangle = 2 \gamma k_B T \delta_{ij} \delta_{ab} \delta(t - t')$, $a, b = x, y, z$ denote the cartesean components, and $T$ is the temperature of the reservoir.  Standard velocity-Verlet algorithm~\cite{allen89,rapaport11} is used to integrate the above Langevin equation in our simulations. 
The atomic forces within the BO-MD framework is computed using the finite temperature generalization of the Hellmann-Feynman theorem $\mathbf f_{i} = - {\rm Tr}(\rho_G \, \partial \mathcal{H} / \partial \mathbf u_i)$. There are two contributions to the atomic forces. The elastic restoring force is  
\begin{eqnarray}
	\mathbf f^l_i = -K_0 \mathbf u_i -K_1\,\sum_j \!^{'} \, \hat{\mathbf e}_{ij}  \cdot  (\mathbf u_i - \mathbf u_j),
\end{eqnarray}
where the summation is restricted to the nearest neighbors. 
The electronic force originates from the $\mathbf u$-dependence of the transfer integral $t_{ij}$, and is given by
\begin{eqnarray}
	\label{eq:f_e}
	\mathbf f^e_{i} = g \sum_{j, \alpha} t_{ij}^{(0)} \hat{\mathbf e}_{ij} \left[ {\rm Tr}(\rho_G \, c^\dagger_{i,\alpha}c^{\;}_{j\alpha} ) + \mbox{c.c.}\right],
\end{eqnarray}
where $\rho_G = \mathcal{P}_G \, \rho^{\rm qp} \, \mathcal{P}_G$ is the Gutzwiller electron density matrix~\cite{sandri13}, $\rho^{\rm qp}$ is the density matrix of quasi-particles, and $\mathcal{P} = \prod_i \mathcal{P}_i$ is the Gutzwiller projector. As in Gutzwiller's original approach, the local projectors $\mathcal{P}_i$ are introduced to reduce the on-site double-occupancy variationally~\cite{gutzwiller63,gutzwiller64,gutzwiller65}. In the modern formulation of the Gutzwiller approximation (GA), these projectors can be conveniently parameterized using the so-called $\Phi$-matrix~\cite{lanata12}, $\mathcal{P}_i \equiv \sum_{\alpha, \beta} \Phi_{i, \alpha\beta} /(\Pi_{i, \beta})^{-1/2} |\alpha\rangle \langle \beta|$, where $\Pi_{i, \beta}$ is the occupation probability of local electron spin-orbital configuration $|\beta\rangle$ for the uncorrelated state. For single-band Hubbard model, the $\Phi$-matrix is diagonal
\begin{eqnarray}
	\Phi_i = \left[ \begin{array}{cccc}
	e_i & 0 & 0 & 0 \\
	0 & p_{i, \uparrow} & 0 & 0 \\
	0 & 0 & p_{i, \downarrow} & 0 \\
	0 & 0 & 0 & d_i
	\end{array} \right],
\end{eqnarray}
and its elements represent the empty, singly occupied (spin up and down), and doubly occupied electron state of the $i$-th atom~\cite{lanata15,lanata17}. As discussed in Sec.~\ref{sec:intro}, they can also be viewed as the coherent-state amplitude of the slave bosons in the Kotliar-Ruckenstein theory, whose saddle-point solution corresponds to the GA.

To obtain the Gutzwiller density matrix $\rho_G$, one needs to self-consistently solve the quasi-particle density matrix $\rho^{\rm qp}$ and the SB amplitudes $\Phi_i$. For a given set of atomic displacements~$\{\mathbf u_i \}$, which defines the tight-binding $t_{ij}$ through Eq.~(\ref{eq:t_ij}), the Gutzwiller solution is obtained by minimizing the electron free energy
\begin{eqnarray}
	\label{eq:free-energy}
	& & \mathcal{F} = {\rm Tr}(\rho^{\rm qp} \mathcal{H}^{\rm qp}) + U \sum_i {\rm Tr}(\Phi^\dagger_i n_{i, \uparrow} n_{i, \downarrow} \Phi_i) \nonumber \\ 
	& &  \quad +  \, T \, {\rm Tr}(\rho^{\rm qp} \ln \rho^{\rm qp}) + T \sum_i {\rm Tr}\left[ \Phi^\dagger_i \Phi^{\;}_i \ln\left(\Phi^\dagger_i \Phi^{\;}_i / \Pi_i\right) \right] \nonumber \\
	 & & \quad + \, \sum_{i, \alpha} \mu_{i,\alpha} \left[ {\rm Tr}(\Phi^\dagger_i \, \hat{n}_{i, \alpha} \Phi^{\;}_i)  - {\rm Tr}(\rho^{\rm qp} \hat{n}_{i, \alpha})\right].
\end{eqnarray}
The first line above denotes the total energy within the Gutzwiller approximation. The two terms in the second line correspond to quasi-particle entropy and the relative entropy between correlated distribution $\Phi^\dagger_i \Phi_i$ and the corresponding uncorrelated $\Pi_i$~\cite{wang10,sandri13}. The atom-dependent Lagrangian multipliers $\mu_{i, \alpha}$ are introduced to enforce the Gutzwiller constraints in the third line, and $\hat{n}_{i\alpha} \equiv c^\dagger_{i\alpha} c^{\;}_{i\alpha}$ is the electron number operator. 

The quasi-particle density matrix which minimizes the free-energy Eq.~(\ref{eq:free-energy}) is given by the Boltzmann distribution $\rho^{\rm qp} = e^{-\beta \mathcal{H}^{\rm qp}} / Z^{\rm qp}$, where $\mathcal{H}^{\rm qp}$ is the renormalized tight-binding Hamiltonian
\begin{eqnarray}
	\label{eq:H_renorm}
	\mathcal{H}^{\rm qp} &=&  \sum_{\langle ij \rangle, \alpha} \left( \mathcal{R}^{\;}_{i,\alpha} \mathcal{R}^*_{j,\alpha}\, t_{ij} c^\dagger_{i,\alpha} c^{\;}_{j, \alpha} + \mbox{h.c.} \right) \nonumber \\
	 & & \qquad - \sum_{i, \alpha} \mu_{i, \alpha}\, c^\dagger_{i, \alpha} c^{\;}_{i, \alpha}, 
\end{eqnarray}
and $t_{ij}$ depends on the displacements through Eq.~(\ref{eq:t_ij}). This Hamiltonian can also be viewed as an effective tight-binding model with renormalized hopping coefficients: $t^{\rm eff}_{ij} = t_{ij} \mathcal{R}^{\,}_{i,\alpha} \mathcal{R}^*_{j,\alpha}$, where the renormalization factors are~\cite{lanata12} 
\begin{eqnarray}	
	\label{eq:R}
	\mathcal{R}_{i, \alpha} = \frac{ {\rm Tr}\left(\Phi^\dagger_i c^\dagger_{i, \alpha} \Phi^{\;}_i c^{\;}_{i, \alpha} \right) }{  \sqrt{n_{i, \alpha} (1 - n_{i, \alpha})} }.
\end{eqnarray}
and $n_{i,\alpha} = {\rm Tr}(\Phi^\dagger_i \, \hat{n}_{i, \alpha} \Phi^{\;}_i)$ is the quasi-particle number at the $i$-th atom.

The minimization of $\mathcal{F}$ with respect to $\Phi_i$, subject to constraint ${\rm Tr}(\Phi^\dagger_i \Phi^{\;}_i) = 1$ can be recast into a nonlinear eigenvalue problem $\mathcal{H}^{\rm SB}_i \Phi_i = \epsilon_i \Phi_i$ for each atom~\cite{lanata12}. Since $\Phi_i$ is diagonal in the case of single-band Hubbard model, we use the same notation $\Phi_i = (e_i, p_{i, \uparrow}, p_{i, \downarrow}, d_i)$ to denote the wavefunction or the coherent state of the slave bosons. The embedded Hamiltonian has the form
\begin{eqnarray}
	\label{eq:H_SB}
	\mathcal{H}^{\rm SB}_i &=& \sum_{\alpha}  \frac{  \left(\Delta_{i, \alpha} \mathcal{M}_{\alpha} + \Delta^*_{i, \alpha} \mathcal{M}^{\dagger}_{\alpha} \right)}{\sqrt{n_{i\alpha} (1-n_{i\alpha})}} + U \,\mathcal{U} \nonumber \\
	  & & + \sum_\alpha  \left( \Lambda_{i\alpha} + \mu_{i\alpha} \right) \mathcal{N}_\alpha + T \Theta_i,
\end{eqnarray}
where $\Delta_{i, \alpha}$ encodes the local electron bonding
\begin{eqnarray}
	\label{eq:Delta}
	\Delta_{i, \alpha} = \sum_j t_{ij} \mathcal{R}^*_{j,\alpha} {\rm Tr}(\rho^{\rm qp}  c^\dagger_{i\alpha} c^{\;}_{j \alpha} ).
\end{eqnarray}
Explicit expressions for the various matrices $\mathcal{M}_\alpha$, $\mathcal{N}_\alpha$, $\mathcal{U}$ and $\Theta_\alpha$, as well as expression for the coefficient $\Lambda_{i, \alpha}$ are given in Appendix~\ref{sec:GA}.
The Hamiltonian $\mathcal{H}^{\rm SB}$ essentially describes the competition between electron bonding favored by the first $\mathcal{M}$ term, and the Coulomb repulsion represented by the second $\mathcal{U}$ term. The Lagrangian multiplier $\mu_{i\alpha}$ appears in the third term acts as a chemical potential in order to satisfy the self-consistent Gutzwiller constraint. Finally, the $\Theta_i$ term describes the entropy contribution~\cite{sandri13}. 

The above two steps, namely solving the quasi-particle density matrix $\rho^{\rm qp}$ and the slave-boson amplitudes $\Phi_i$, have to be iterated many times until self-consistency is reached at every MD time-step. In particular, the Gutzwiller constraint 
\begin{eqnarray}
	{\rm Tr}(\Phi^\dagger_i \Phi^{\;}_i \hat{n}_{i, \sigma})  = {\rm Tr}(\rho^{\rm qp} \hat{n}_{i, \alpha}) = n_{i,\alpha},
\end{eqnarray}
has to be satisfied. 
The bottleneck of the MD simulation here is solving the renormalized tight-binding Hamiltonian, which is required at every Gutzwiller iteration in every time-step. In our simulations, we adopt the gradient-based probing technique combined with the kernel polynomial method (KPM)~\cite{barros13,wang18,chern18} to solve the renormalized tight-binding model. This allows for an efficient estimation of the electron density matrix $\rho^{\rm qp}$ at a computational cost that scales linearly with system size. All simulations discussed below were performed on triangular lattice of unprecedented size of $N = 150^2$ sites. 

In this work we focus on the intrinsic Mott transition in the absence of magnetic ordering. We consider only non-magnetic Gutzwiller solutions, $p_{i, \uparrow} = p_{i, \downarrow} = p_i$, and $n_{i, \uparrow} = n_{i, \downarrow} = n_i$, in all simulations discussed below. 
The following model parameters are used. $t^{(0)}_{\rm nn1} = -1$, $g = 0.55$, $K_0 = K_1 = 0.35$, $m = 2$, $\gamma = 0.2$, and the time step $\Delta t = 0.02$. The number of the Chebyshev polynomials $M = 2000$ is used for the KPM calculation. In order to assist transitions between the local minima of the local order parameter $\Phi_i$, a small noise of the order of $\delta\Phi_i = 0.1$ is added to the slave-boson amplitudes obtained from the previous time-step at the beginning of the new Gutzwiller iteration.


\section{Thermal Mott transition}
\label{sec:thermal-mott}

\begin{figure}
\includegraphics[width=0.99\columnwidth]{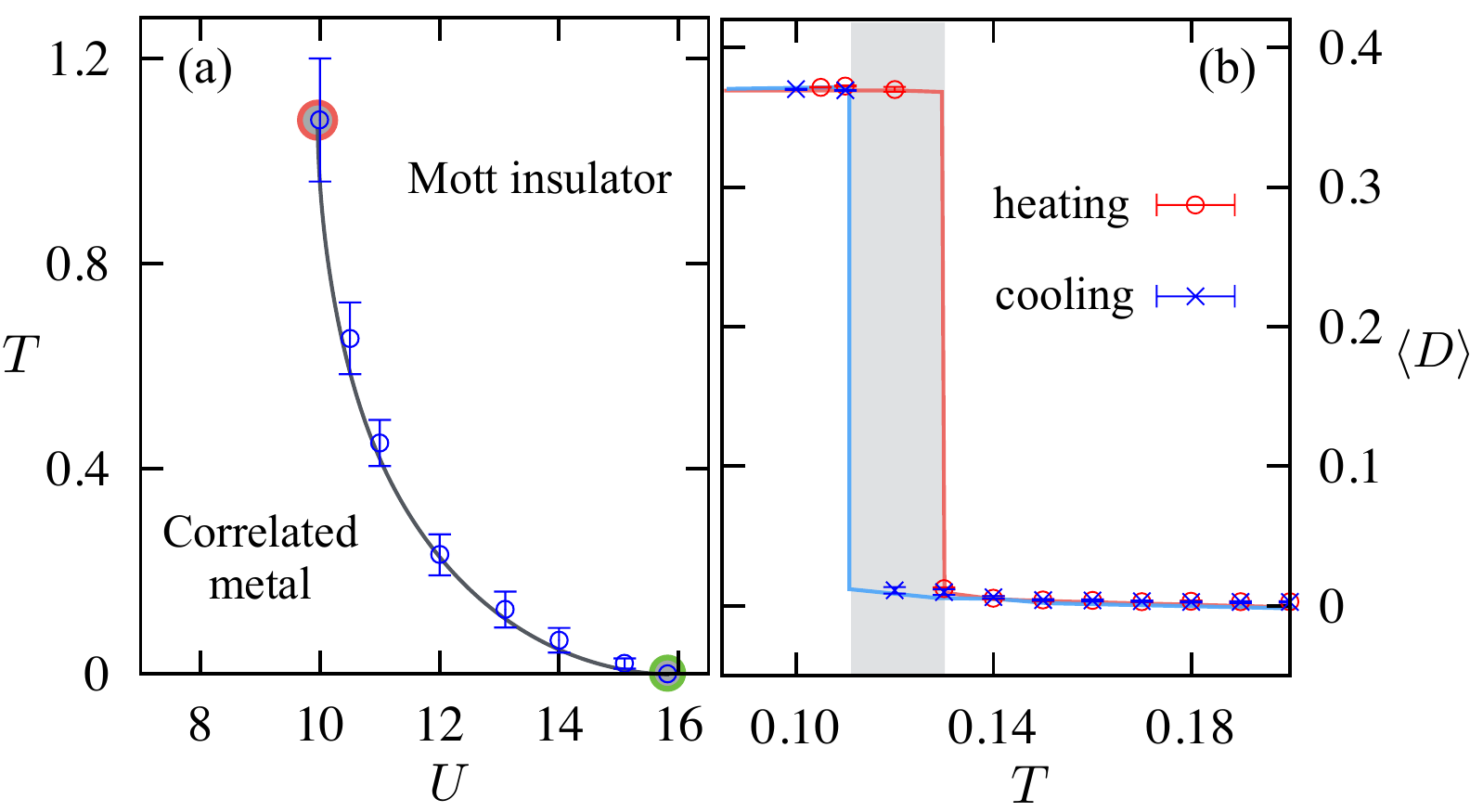}
\caption{(Color online)  
\label{fig:phase-diagram} (a) Schematic phase diagram of the half-filled Peierls-Hubbard model on the triangular lattice. The phase boundary is estimated from the Gutzwiller MD simulations. The nearest-neighbor hopping $t^{(0)}_{1} = 1$ serves as the units for energy and temperature. The critical line $U_c(T)$ ends at $T = 0$ and $T_c \approx 0.9$. (b) The spatially averaged double-occupancy $\langle D \rangle = \sum_i^N D_i/N$ versus the temperature with the Hubbard repulsion set at $U = 13.1$. The shaded area marked the coexistence region of the phase transition.
}
\end{figure}

A schematic phase diagram of the half-filled Peierls-Hubbard model is shown in Fig.~\ref{fig:phase-diagram}(a) based on the Gutzwiller MD simulations. Similar phase diagrams are obtained from the finite-temperature Gutzwiller calculation~\cite{wang10,mezio17} as well as DMFT~\cite{georges96,bulla01,liebsch09}. The first-order MIT line $U_c(T)$ ends at two critical points at $T = 0$ and a finite $T_c > 0$.   The first analytical theory of the zero-temperature MIT was proposed by Brinkman and Rice using the Gutzwiller method~\cite{brinkman70}. In the Brinkman-Rice scenario, the double-occupancy decreases linearly with increasing $U$ according to: $D(U) = \frac{1}{4}(1 - U/U_c)$, where the critical $U_c = 8 |E_{\rm band}|$ and $E_{\rm band}$ is the electron band energy. It is worth noting that the quasi-particle behavior in the Mott transition is well captured, at least qualitatively, by the Brinkman-Rice theory, although it cannot describe the incoherent electronic excitations and the Hubbard bands. 

At finite temperatures, the Mott transition becomes first-order with the double-occupancy showing a discontinuity $\Delta D$ at the transition point. This is demonstrated in Fig.~\ref{fig:phase-diagram}(b) which shows the temperature dependence of the spatially averaged double-occupancy $\langle D \rangle$ serving as an order parameter of the Mott transition. In addition to a clear discontinuity in $\langle D \rangle$, the first-order nature of the MIT also manifests itself in the hysteresis loop from the heating and cooling curves.

Here we consider the evolution of the Hubbard model subject to a temperature quench. For simplicity, we assume the temperature of the system is controlled by a reservoir which only interacts with the lattice through the stochastic forces $\bm\xi_i(t)$ in the Langevin equation~(\ref{eq:langevin}). The system is initially equilibrated at a temperature $T_i$ in the metallic phase. At $t = 0$, the reservoir temperature is suddenly increased to a higher temperature $T_f$ across the phase boundary. Interestingly, due to the negative slope of the phase boundary $dU_c/dT < 0$~\cite{note-dUdT}, such a thermal quench drives the system from the correlated metal to a Mott insulator. 
Within the BO approximation, the electrons are assumed to quickly reach equilibrium at the instantaneous lattice temperature, i.e. $k_B T_e(t) = m \langle \dot{\mathbf u}(t)^2 \rangle /3$. The electron temperature thus increases continuously from $T_i$ to $T_f$ as the lattice is heated up by the reservoir.

\begin{figure}
\includegraphics[width=0.95\columnwidth]{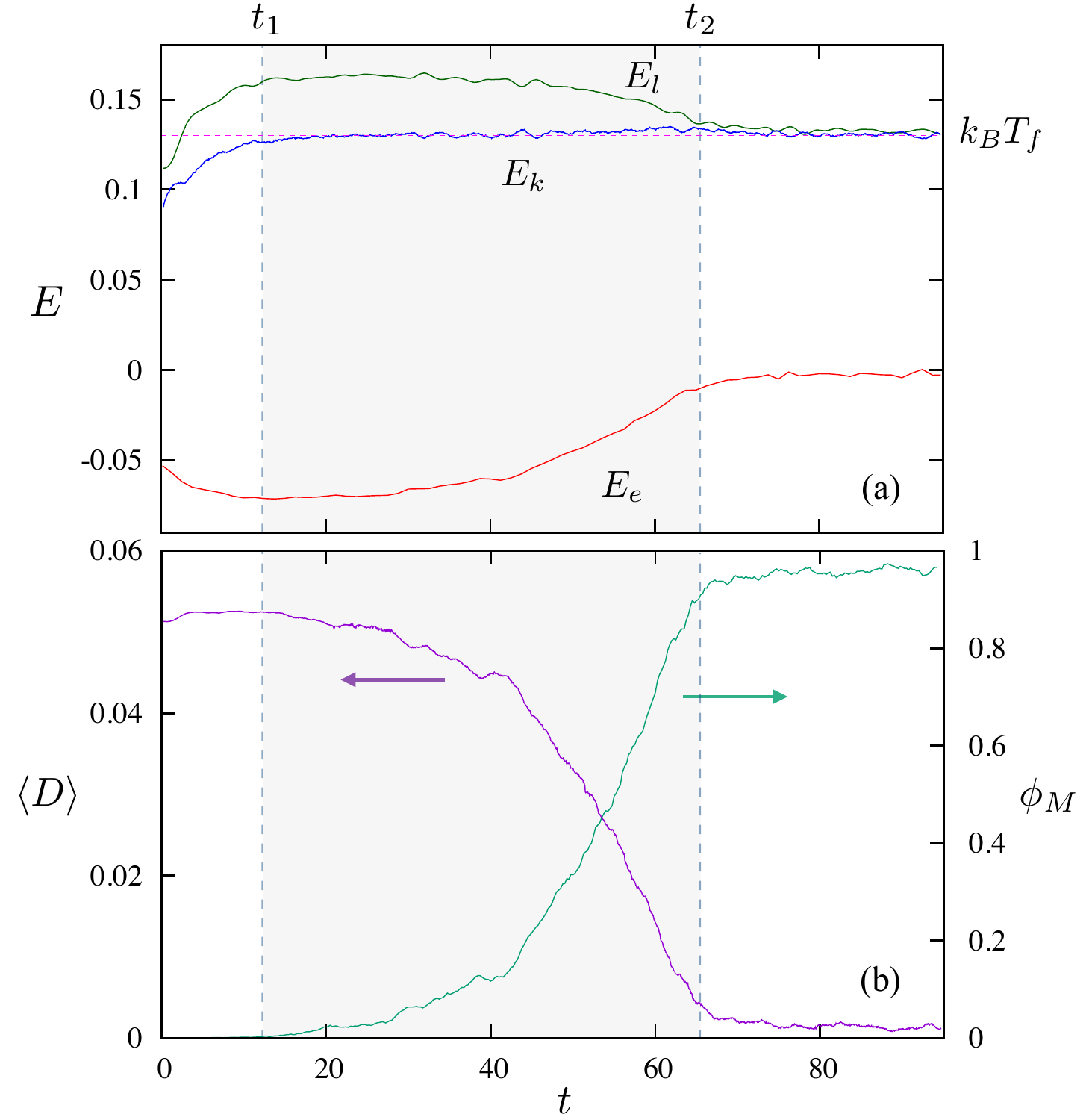}
\caption{(Color online)  
\label{fig:trace} (a) The various energy densities versus time in a MIT driven by a temperature quench. Here $E_e$, $E_l$, and $E_k$ denote the electronic, lattice (elastic), and atomic kinetic energies, respectively. The initial and final temperatures are $T_i = 0.09$ and $T_f = 0.13$, respectively. The (red) horizontal dashed line corresponds to $k_B T_f$. (b) The averaged double occupation $\langle D \rangle$ and the volume fraction of the Mott phase $\phi_M$ versus time. The shaded area bounded between $t_1$ and $t_2$ indicates the region where most of the phase transformation takes place.
}
\end{figure}

The time dependence of the various energy densities is shown in Fig.~\ref{fig:trace}(a) for a quench from $T_i = 0.09$ to $T_f = 0.13$. The electronic energy $E_{\rm elec}$ includes both the hopping and Hubbard term. The lattice energy $E_l$ comes from the elastic $K$ terms in the Hamiltonian. The atomic kinetic energy density is $E_k = m \sum_i |\dot{\mathbf u}|^2 / 2N$.  The phase transformation mostly takes place in the temporal regime $t_1 < t < t_2$, marked by the shaded area. Immediately after the quench, the lattice temperature gradually increases toward the final temperature $T_f$. The enhanced thermal fluctuation of displacements in this initial period ($t < t_1$) result in an increase in elastic energy $E_l$. The enlarged fluctuations of $\langle |\mathbf u_i|^2 \rangle$ also lead to an enhanced variance of the nearest-neighbor hopping $t_{ij}$, which in turn gives rise to a slightly reduced electron energy. However, this relatively short transient phase does not play a significant role in terms of phase transformation.

\begin{figure}[b]
\includegraphics[width=0.99\columnwidth]{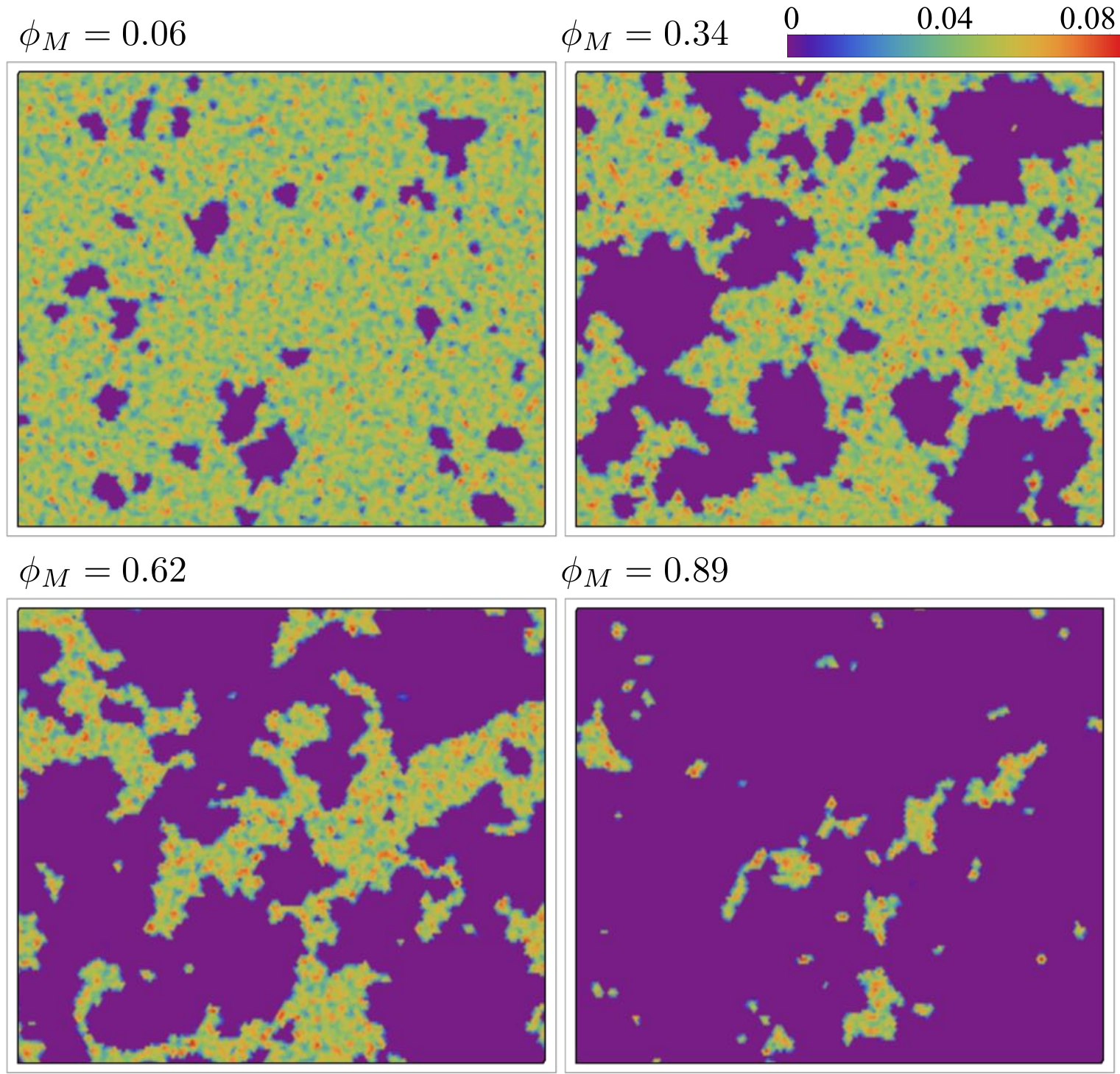}
\caption{(Color online)  
\label{fig:snapshots} Snapshots of the on-site double occupancy $D(\mathbf r_i) = \langle n_{i, \uparrow} n_{i, \downarrow} \rangle$ at four different Mott volume fractions, corresponding to $t = 30$, 50, 58, and 65, during the thermal driven MIT. See Fig.~\ref{fig:trace}(b) for the corresponding time dependence of the Mott volume fraction and the average double-occupancy.  
}
\end{figure}


As discussed in Sec.~\ref{sec:intro}, the double occupancy is one of the order-parameters for characterizing the Mott transitions. Another important quantity is the volume fraction of Mott phase $\phi_M$, which is often used in the study of phase transformation kinetics. The time dependence of these two quantities during the thermal Mott transition is shown in Fig.~\ref{fig:trace}(b); here $\langle D \rangle = \sum_i D_i / N$ is the spatial average of the local double occupation. Numerically, we label a given site as the Mott insulator if its double-occupancy $D_i < 10^{-8}$. It is worth noting that the transformed volume fraction $\phi_M$ is negligible in the initial period when $t < t_1$. In fact, the average $\langle D \rangle$ even slightly increases due to the reduced electron energy. 

After this initial period, the system is trapped in a meta-stable metallic state with a temperature $T \approx T_f$.
The transformation to the Mott insulator mainly takes place in the temporal window $[t_1, t_2]$, in which the Mott volume fraction $\phi_M$ gradually increases from zero to its equilibrium value at $T_f$ while the average $\langle D \rangle$ slowly decays to zero, see Fig.~\ref{fig:trace}(b).
Once the system reaches the new equilibrium at $T_f$ for both the lattice and electrons, the electron binding energy is renormalized to almost zero. This also indicates a vanishing electronic force $\mathbf f_i^e \approx 0$; see Eq.~(\ref{eq:f_e}). Consequently, the lattice system becomes essentially a collection of free harmonic oscillator, and both energy densities $E_l$ and $E_k$ approach~$k_B T_f$ consistent with the equipartition theorem.

\section{Nucleation and growth}
\label{sec:nucleation}

\begin{figure}
\includegraphics[width=0.99\columnwidth]{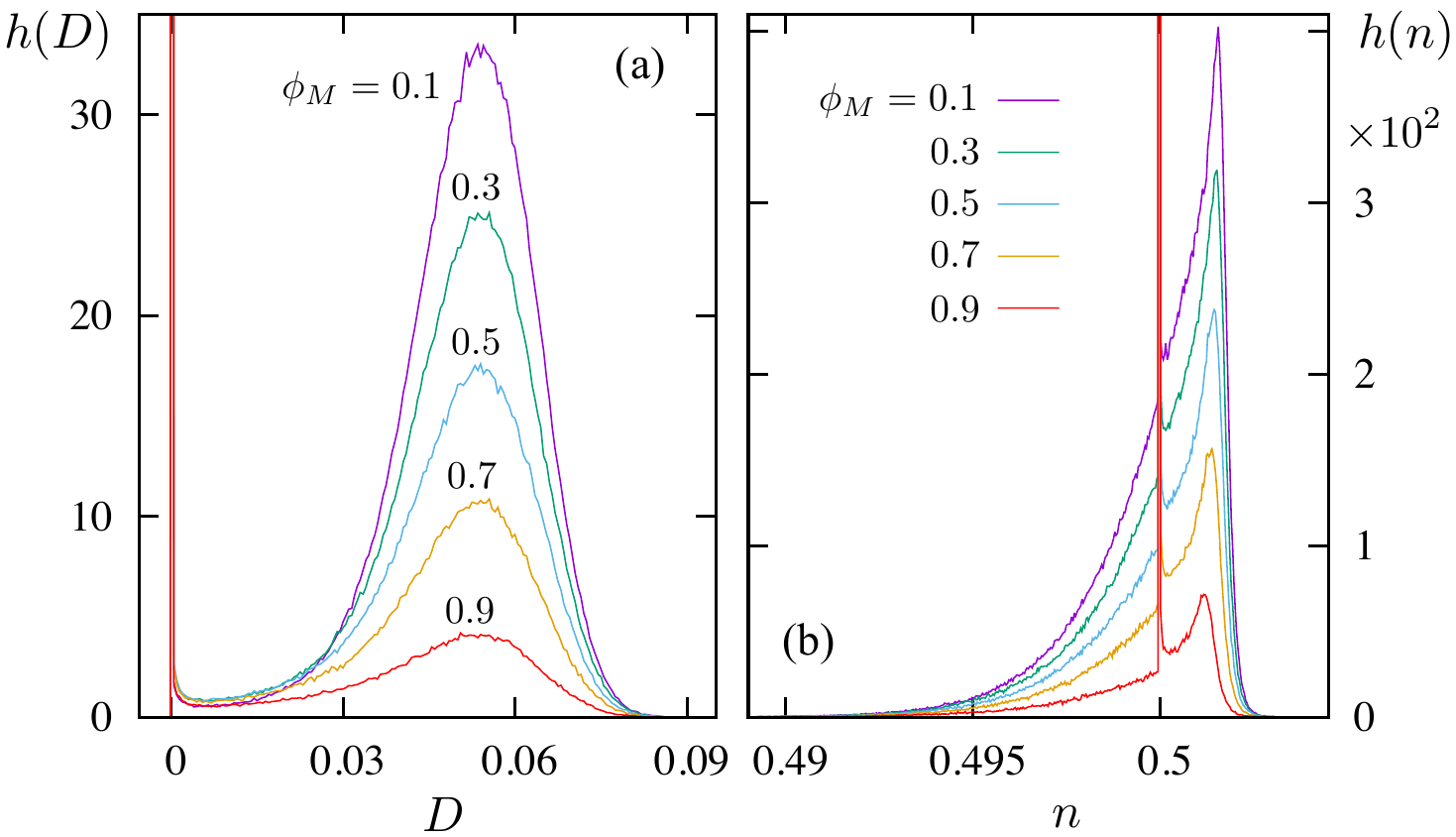}
\caption{(Color online)  
\label{fig:hist_db} Computed histogram of (a) double-occupancy and (b) on-site electron density at varying Mott volume fraction for the temperature-quench simulations of Fig.~\ref{fig:trace}.
}
\end{figure}

To further investigate the nature of the phase transformation, Fig.~\ref{fig:snapshots} shows the spatial profile of the local double-occupancy $D(\mathbf r_i) = \langle n_{i, \uparrow} n_{i, \downarrow} \rangle$ at four different times during the relaxation. These snapshots clearly show that the transition to the new phase is initiated by the nucleation of Mott insulating puddles, i.e. regions with vanishing double-occupancy, in a metallic background. Importantly, the double-occupation in the metallic regions remains roughly the same. This observation also means that the gradual decrease of the average $\langle D \rangle$ in Fig.~\ref{fig:trace}(b) results mostly from the increasing volume fraction of the Mott clusters in a heterogeneous electronic state, instead of the decreasing double-occupancy of a homogeneous state. 

This scenario is further confirmed by the histogram or probability distribution function~$h(D)$ of the double-occupancy, shown in Fig.~\ref{fig:hist_db}(a), at various Mott volume fractions. The distribution function consists mainly of a delta-function at $D = 0$ corresponding to the Mott clusters, and a broad hump centered at $D \approx 0.055$, which remains roughly the same throughout the transition. This result is consistent with the picture of two coexisting phases in the intermediate state. As the phase transformation progresses, the delta-peak keeps increasing at the expense of the finite-width distribution at finite~$D$. Interestingly, the electron density $n(\mathbf r_i)$ exhibits a rather uniform distribution, as demonstrated by its histogram $h(n)$ shown in Fig.~\ref{fig:hist_db}(b). The on-site electron density $n_i$ is very narrowly distributed around the half-filling with a width $\delta n \lesssim 0.01$, and is dominated by a huge delta-peak at $n = 0.5$, with contributions from both the metallic and Mott regions.

\begin{figure}
\includegraphics[width=0.95\columnwidth]{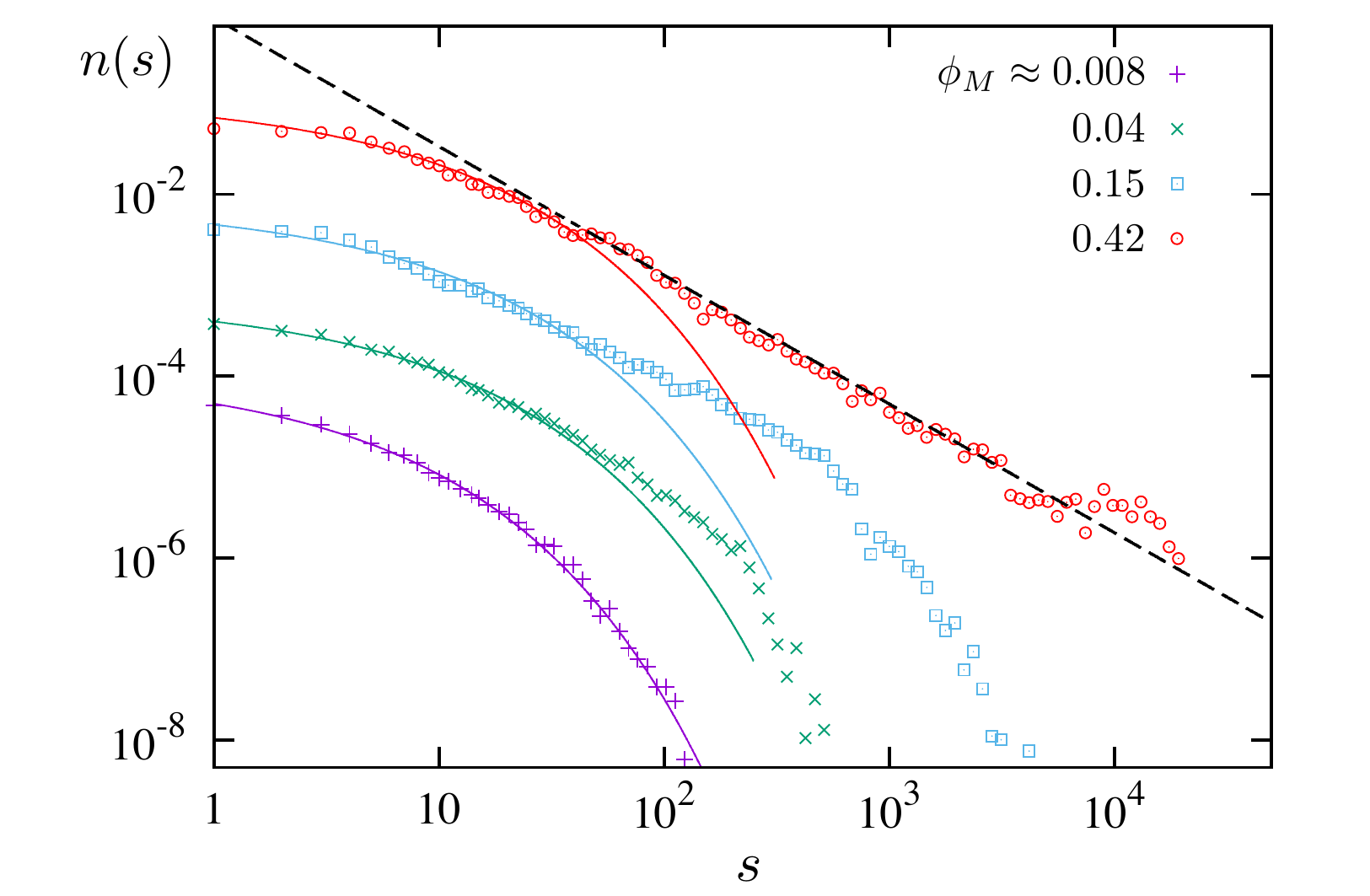}
\caption{(Color online)  
\label{fig:histogram} The probability distribution function $n(s)$ of the size of Mott clusters at different stages, characterized by the volume fraction of the Mott clusters, of the phase transformation. The curves are shifted vertically for clarity. The corresponding curve of volume fraction $\phi_M$ versus time is shown in Fig.~\ref{fig:trace}(b). Each curve is obtained by average over a small time interval $[t-\Delta t, t+\Delta t]$ with $\Delta t = 2.5$, and over 15 independent simulations. The solid lines are the fit using Eq.~(\ref{eq:ns_exp}). The dashed line corresponds to power law $s^{-\tau'}$ with $\tau' = 1.42$. The estimated error of the exponent is $\pm 0.06$.
}
\end{figure}

During the initial phase transformation, both the number and size of the Mott clusters increase with time. After a while, these Mott clusters start to merge through a process akin to the percolation transition. Of central interest here is the evolution of the cluster-size distribution $n(s, t)$, which is shown in Fig.~\ref{fig:histogram} for different stages of the phase transformation. For small clusters, it is expected that their distribution is mainly determined by the interface energy which can be expressed as $\varepsilon(A) = \sigma A$, where $\sigma$ is the surface tension and $A \sim s^{(d-1)/d}$ is the surface area of nuclei with size $s$. In two dimensions their distribution, given by the Boltzmann probability, is then
\begin{eqnarray}
	\label{eq:ns_exp}
	n(s, t) \sim \exp(- \beta \sigma_{\rm eff} \sqrt{s}),  
\end{eqnarray}
where $\sigma_{\rm eff}$ is an effective surface tension. As shown in Fig.~\ref{fig:histogram}, Eq.~(\ref{eq:ns_exp}) indeed fits the numerical distributions very well for small $s$, particularly at the initial stage of the phase transformation. The effective surface tension obtained from the fitting is $\sigma_{\rm eff} \approx 0.1$.

Significant deviation from Eq.~(\ref{eq:ns_exp}) starts to occur at cluster size $s_c \approx 100$, which provides an estimate for the so-called critical size. The concept of critical nuclei was first discussed in the classical droplet model~\cite{gunton83}. In this picture, the decay of a meta-stable state is initiated by a collection of independent fluctuating droplets. Their quasi-equilibrium distribution is again given by the Boltzmann probability $n_0(s) \propto \exp[- \beta \varepsilon(s)]$, where $\varepsilon(s)$ is the minimum work required to form a cluster of size $s$. In general $d$ dimensions, this minimum work is
\begin{eqnarray}
	\varepsilon(s) = \sigma s^{(d-1)/d} + \Delta f\, s.
\end{eqnarray}
The first term represents the interface energy as discussed above, while the second term denotes volume energy of the droplets with $\Delta f$ being the free-energy difference between the two competing phases. For $\Delta f > 0$, the droplets correspond to thermodynamically unfavorable phase, and $n_0(s)$ describes the fluctuations of these small domains. On the other hand, for $\Delta f < 0$, as in a quench into meta-stable states, $\varepsilon(s)$ describes the competition between the surface tension and the volume energy gain. In this case,  the distribution $n_0(s)$ decreases as a function of $s$ for small clusters until it reaches the minimum at the critical size $s_c \sim (\sigma/ |\Delta f|)^d$. For $s > s_c$, the probability $n_0(s)$ then grows exponentially with increasing cluster size. These super-critical clusters are thus seeds of the nucleation process. The maximum of the work function $\varepsilon_{c} = \varepsilon(s_c) \sim \sigma^d/|\Delta f|^{d-1}$ represents an energy barrier to nucleate a critical cluster.

\begin{figure}[b]
\includegraphics[width=0.99\columnwidth]{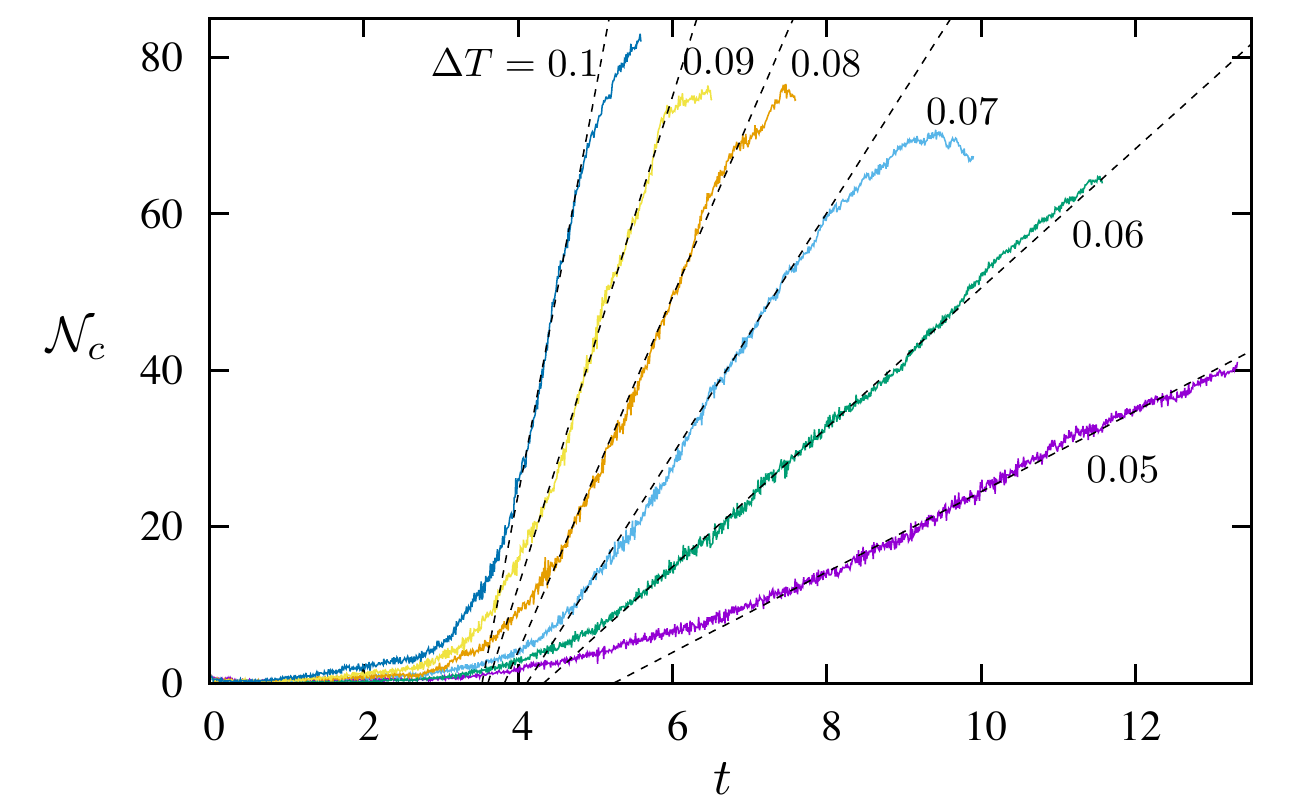}
\caption{(Color online)  
\label{fig:Nc} The total number of nuclei $\mathcal{N}_c$ as a function of time for various $\Delta T = T_f - T_i$ from the Gutzwiller MD simulations. The initial temperature $T_i = 0.09$. The dashed lines are fit to the linear equation in Eq.~(\ref{eq:linear}).
}
\end{figure}

Once the phase transformation is initiated by these seed clusters, the quasi-equilibrium $n_0(s)$ no longer describes the time-dependent cluster distribution. Numerical methods are often required to compute the evolution of distribution function $n(s, t)$. As Fig.~\ref{fig:histogram} shows, the distribution function extends beyond the estimated critical size with an increasingly longer tail. Interestingly, as the volume fraction approaches $\phi_M^* \approx 0.42$, the cluster-size distribution exhibits a power-law tail
\begin{eqnarray}
	\label{eq:tau}
	n(s, t^*) \sim 1/s^{\tau'}, 
\end{eqnarray}
with an exponent $\tau' \approx 1.42$. In percolation transitions, the size of the percolating clusters follows a similar power-law distribution, $n(s; p \to p_c) \sim s^{-\tau}$, when the percolation probability $p$ approaches the critical value $p_c$. Here $\tau$ is called the Fisher's exponent, and its numerical value is $\tau = 187/91 \approx 2.055$ for 2D percolation~\cite{stauffer94}. Despite the similarity, it is worth noting that the nature of these two power-law behaviors is very different. The distribution function $n(s, p)$ in the case of percolation is stationary and describes the static structure of the percolating clusters. The emergence of the power-law characterized by Fisher's exponent $\tau$ also indicates a divergent length scale $\xi \sim (p - p_c)^{-\nu}$ related to the underlying critical point~\cite{stauffer94}. On the other hand, the power-law in Eq.~(\ref{eq:tau}) describes a {\em transient} distribution of the Mott clusters during a first-order phase transformation. It does not necessarily imply a divergent length scale, also see discussion in Sec.~\ref{sec:structure}.
On the other hand, the integrated distribution $\mathcal{D}(s) = \int_0^{t^*} n(s, t) dt \sim 1/s^{\tau}$ from our simulations also exhibits a power-law tail with $\tau \approx 2.03$, although it is less pronounced due to the limitation of system sizes. This power-law tail might be related to the Ising critical end point of the Mott transition, where the distribution of the geometrical spin clusters is described by a similar power-law behavior with $\tau =  379/187 \approx 2.027$~\cite{janke05}. A more detailed study of the phase coexistence regime is required to clarify this connection.



We next consider the time dependence of the total number of nuclei, which is related to the distribution function via the integral $\mathcal{N}_c(t) = \int_0^\infty n(s, t) \, ds$. Fig.~\ref{fig:Nc}~shows the numerically computed $\mathcal{N}_c$ versus time for varying depths of the temperature quench $\Delta T = T_f - T_i$. Here we only show the early stages of the nucleation, in which two distinct kinetic behaviors can be seen from the figure: (i)~the incubation period with negligible number of clusters, and (ii) the constant nucleation regime in which the total number of nuclei increases linearly with time. To understand these behaviors, we briefly discuss the classical nucleation theory first developed by Becker and D\"oring~\cite{becker35}, Zedolvich~\cite{zeldovich43}, and Frenkel~\cite{frenkel46}. The distribution function $n(s, t)$ is governed by the Fokker-Planck equation~\cite{gunton83} 
\begin{eqnarray}
	\label{eq:FK}
	\frac{\partial n}{\partial t }= -\frac{\partial J}{\partial s} =  \frac{\partial }{\partial s}\left[ R \, n_0\, \frac{\partial }{\partial s} \left(\frac{n}{n_0} \right) \right],
\end{eqnarray} 
where $J$ is the flux of nuclei in the cluster-size space, with its definition given in the second equality, and $R = R(s)$ is a size-dependent diffusion coefficient. This theory can be viewed as describing a random-walk process where the ``position" of the walker corresponds to the cluster-size $s$. The expression for the flux $J$ in Eq.~(\ref{eq:FK}) can be understood as follows. In general, there are two contributions to the flux: $J = -R \,\partial n / \partial s + A n$, where the first term describes diffusion process, while the second term represents particle drift. The drift-velocity coefficient $A$ is related to $R$ by the Einstein relation which originates from the condition that the flux vanishes for the equilibrium distribution $n(s, t) = n_0(s)$. Substituting $n_0(s)$ into the equation $J = 0$ gives $A n_0 = R\, d n_0 / ds$, which after some algebra corresponds to the flux expression in Eq.~(\ref{eq:FK}).

\begin{figure}[t]
\includegraphics[width=0.99\columnwidth]{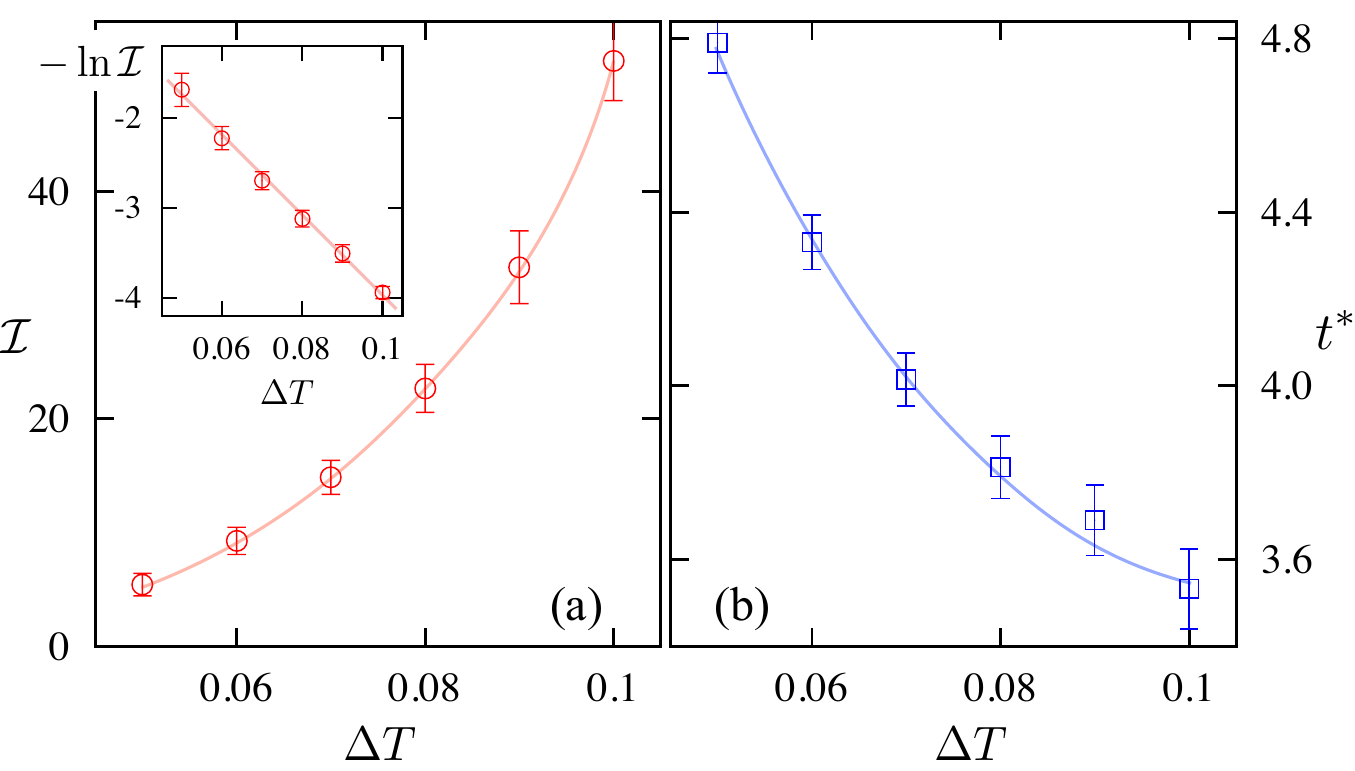}
\caption{(Color online)  
\label{fig:DT_dep} (a) The nucleation rate $\mathcal{I}$ and (b) induction time $t^*$ as functions of the quench depth $\Delta T$ extracted from the curves in Fig.~\ref{fig:Nc}. The inset in panel (a) shows the negative logarithm of $\mathcal{I}$ versus $\Delta T$. The lines are guide for~eye.
}
\end{figure}

The equilibrium distribution $n_0(s)$ of course describes a steady-state solution of the Fokker-Planck equation. 
The regime with a constant nucleation rate, on the other hand, corresponds to a nonequilibrium steady-state solution with $J(s, t) = \mathcal{I} = $ constant, subject to the so-called ``sink and source" boundary conditions: $n(s \to \infty, t) = 0$ and $n(s, t) \to n_0(s)$ for $s \ll s_c$~\cite{becker35}. In this case, the total number of clusters satisfies the simple differential equation $d\mathcal{N}_c/dt = -\int_0^\infty (\partial J/\partial s) \, ds = \mathcal{I}$, which can be readily integrated to yield
\begin{eqnarray}
	\label{eq:linear}
	\mathcal{N}_c(t) =  \mathcal{I} (t - t^*),
\end{eqnarray}
where we have introduced the so-called induction time~$t^*$ as the constant of integration. Our Gutzwiller MD simulations clearly show such a constant nucleation regime at early stage of the phase transformation; see Fig.~\ref{fig:Nc}. The numerically extracted nucleation rate $\mathcal{I}$ and induction time $t^*$ are shown in  Fig.~\ref{fig:DT_dep} as a function of the quench depth $\Delta T$. As expected, increasing the depth of the quench enhances the nucleation rate, while shortening the incubation time for nucleation. The constant flux $\mathcal{I}$ can be computed from Eq.~(\ref{eq:FK}),
\begin{eqnarray}
	\mathcal{I} = \left( \int_0^\infty \frac{ds}{R\,n_0} \right)^{-1} \approx \mathcal{I}_0 \, \exp \! \left(\frac{-\varepsilon_c}{k_B T_f} \right).
\end{eqnarray}
Here the saddle-point approximation has been used to compute the integral, $\varepsilon_c = \sigma^2/2 |\Delta f|$ is the nucleation barrier discussed previously, and the prefactor $\mathcal{I}_0$ is related to the curvature at the work function maximum~\cite{gunton83}. In the inset of Fig.~\ref{fig:DT_dep}(a), we plot the negative logarithm of the nucleation rate as a function of $\Delta T$. Since $\varepsilon_c = - k_B T_f \ln (\mathcal{I}/\mathcal{I}_0)$, our result shows that the energy barrier decreases roughly linearly with the quench~depth.

\section{Interface-controlled domain growth}
\label{sec:interface}

\begin{figure}
\includegraphics[width=0.99\columnwidth]{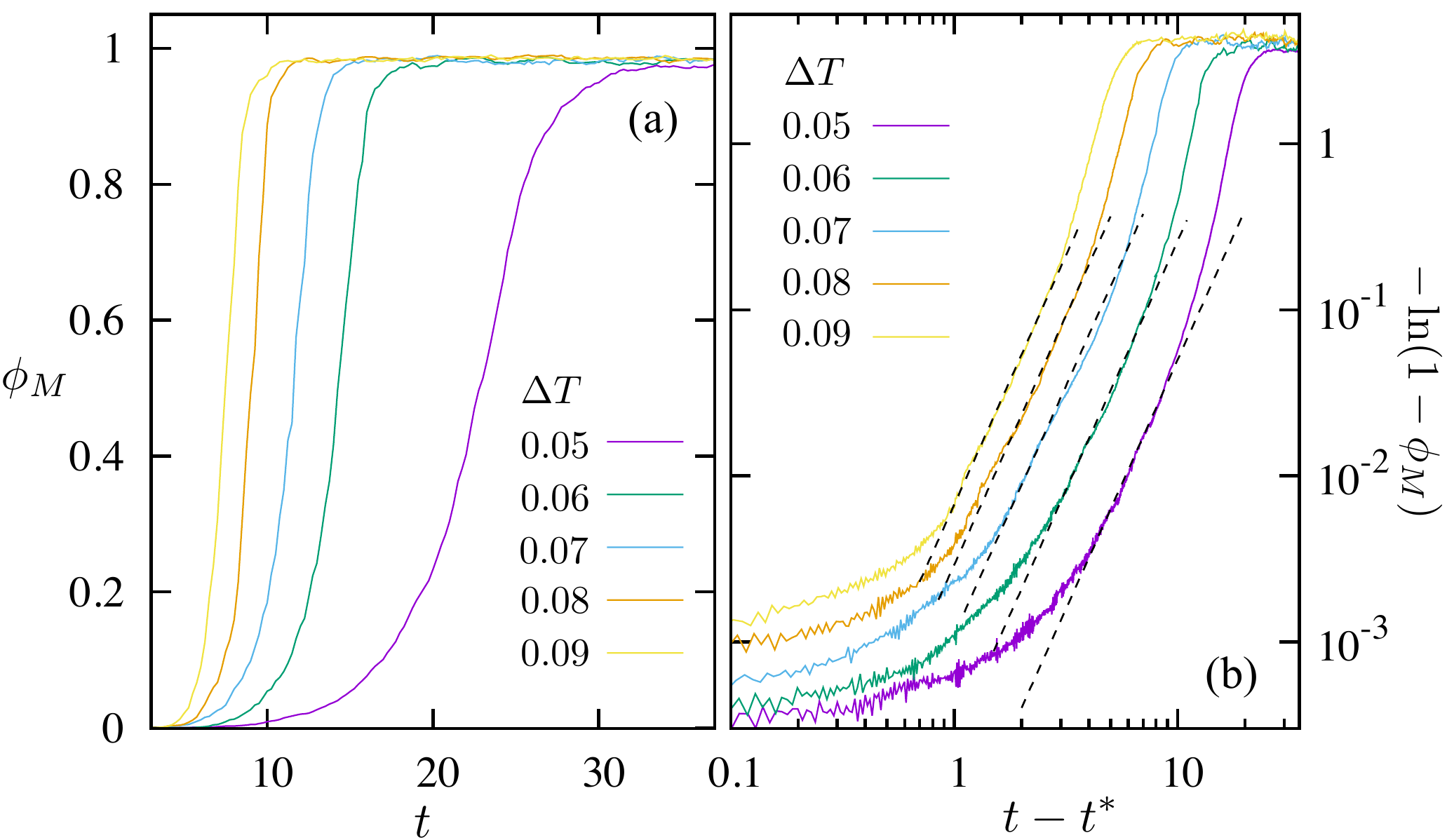}
\caption{(Color online)  
\label{fig:phi_M} (a) The area fraction $\phi_M$ of the Mott phase versus time for varying quench depth $\Delta T = T_f - T_i$, here the initial temperature $T_i = 0.09$. (b) The log-log plot of $-\ln(1-\phi_M)$ versus the shifted time $t - t^*$. The dashed lines correspond to the Avrami equation with an exponent $n = 3$.
}
\end{figure}

After the initial nucleation period, the Mott clusters start to occupy contiguous areas and their growth is hindered. A quantity of central interest here is the volume fraction of the transformed phase. In our case, we are interested in the fraction $\phi_M$  of the Mott phase, which is already introduced in Fig.~\ref{fig:trace}(b) and is shown in Fig.~\ref{fig:phi_M}(a) as a function of time for varying depths $\Delta T$ of the temperature quench.
The standard model for the time dependence of the transformed volume fraction was developed by Kolmogorov, Johnson, Mehl, and Avrami (KJMA model) in the thirties~\cite{kolmogorov37,avrami39,avrami40,johnson39}. The KJMA model is a mean-field type theory for treating the impingement of clusters. To account for this effect,  KJMA introduced the concept of extended volume $V^e$, which is the volume occupied by the new phase {\em if} they can grow without being hindered by the presence of other nuclei. 
Let $dV_e = d\phi_e \mathcal{V}$ be the increase of the extended volume within a small time interval $dt$, and $\mathcal{V}$ is the fixed total volume of the system. As a fraction $\phi$ of the system has been transformed to the new phase, the free-space available for cluster growth is a fraction $f = 1- \phi$ of the total volume. Consequently, the real volume increase  $dV$ that is available for the  growing nuclei is expected to be also the same fraction $ f = 1 - \phi$ of the extended volume, i.e. $dV/dV^e = d\phi/d\phi^e = f$. Integrating this equation leads to a relation between the fraction of the actual transformed volume and that of the extended one
\begin{eqnarray}
	\label{eq:avrami}
	\phi(t) = 1 - \exp[-\phi^e(t)] = 1 - \exp\left( - K t^n \right).
\end{eqnarray}
In the second expression, we have assumed a power-law growth of the extended volume and $K$ is a pre-factor proportional to the nucleation rate. This simple expression, also called the Avrami equation, has been widely used in the analysis of phase transformation experiments. It has been suggested that $n$ can be partitioned as $n = \alpha + d p$~\cite{christian75}, where $a$ depends on the nucleation rate ($\alpha = 0$ for existing nuclei, and $\alpha = 1$ for constant nucleation),  $d$ is the spatial dimensions, and $p$ is related to the growth mechanism. Notably, $p = 1$ for interface-controlled growth, and $p = 1/2$ for diffusion controlled process. 


From the Gutzwiller MD simulations, we can extract the exponent $n$ by plotting the effective extended fraction $ \phi^e_M = -\ln(1 - \phi_M)$ versus the simulation time, as shown in Fig.~\ref{fig:phi_M}(b). Since the Avrami equation is obtained under the assumption of constant cluster growth, the time here is shifted by the induction time $t^*$ obtained in Fig.~\ref{fig:DT_dep}(b). The log-log plot clearly shows a linear regime with slope $n = 3$ for small $t$. This result is consistent with the standard interface-controlled ($p = 1$) homogeneous nucleation ($a=1$), giving rise to $n = d + 1 = 3$ in our case. In general, the evolution of the extended volume fraction can be expressed as $\phi^e(t) \propto \int_0^t  \langle {s}(t; t') \rangle\, \mathcal{I}(t') \, dt'$, where $\mathcal{I}(t)$ is a time-dependent nucleation rate and $\langle {s}(t; t') \rangle$ is the size of nucleus at time $t$ that was nucleated at $t'$. For constant nucleation rate, as demonstrated in Sec.~\ref{sec:nucleation} for the early stage of the transformation, and assuming a time-invariant cluster growth, we have $\phi^e(t) \propto \mathcal{I} \int_0^t \langle {s}(t - t') \rangle dt'$.

\begin{figure}
\includegraphics[width=0.98\columnwidth]{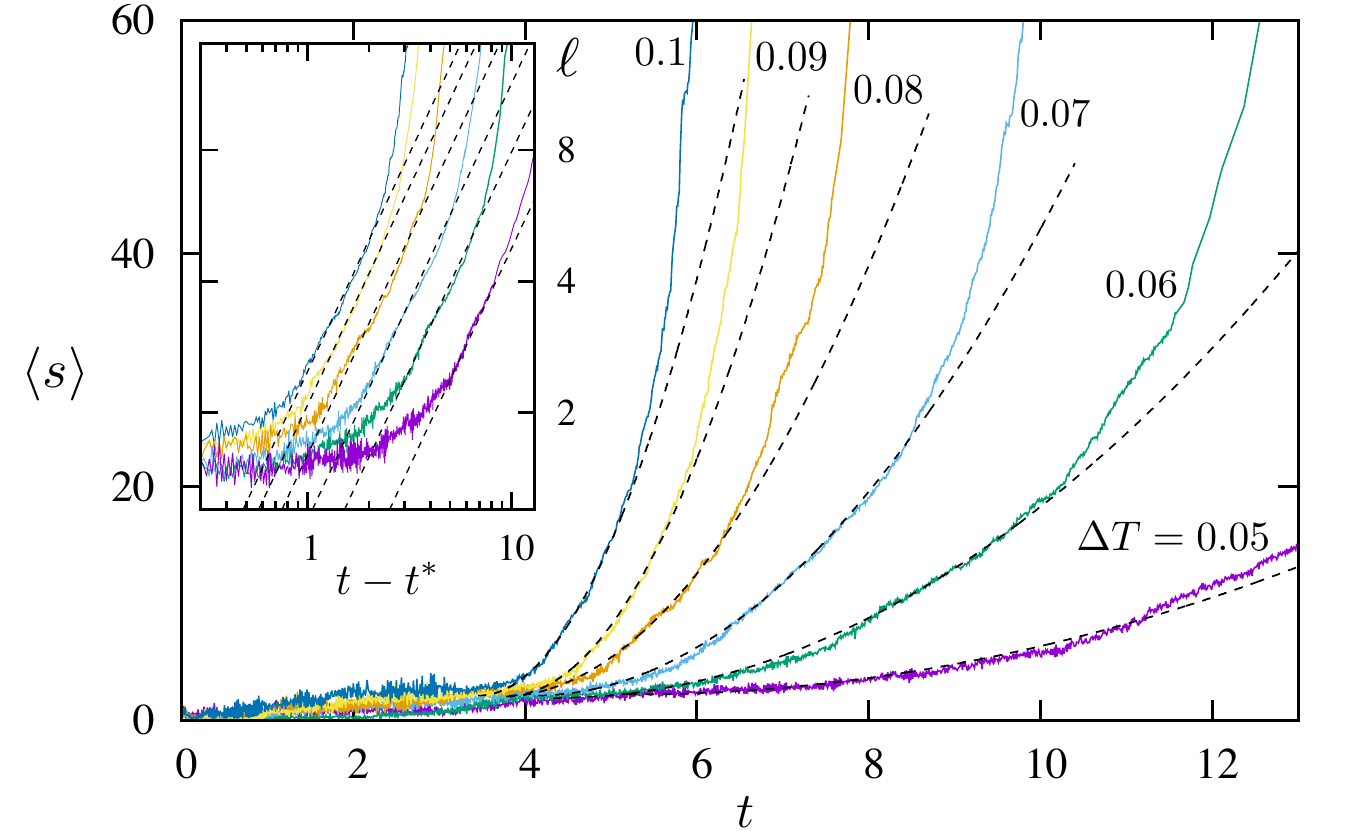}
\caption{(Color online)  
\label{fig:size} The average size of nuclei $\overline{s}$ as a function of time for varying depths of the temperature quench. The dashed lines are the fit with parabolic functions. The inset shows the log-log plot of average linear size of clusters ${\ell} = \langle {s} \rangle^{1/2}$ versus the shifted time $t - t^*$. The dashed lines correspond to linear function ${\ell} \sim (t - t^*)$. 
}
\end{figure}

In the case of interface-controlled mechanism, the growth rate of a cluster of size $s$ is proportional to its surface ``area" $A \sim s^{(d-1)/d}$, i.e. $ds/dt \sim s^{(d-1)/d}$, from which one obtains a growth law
\begin{eqnarray}
	\langle {s} \rangle \sim t^d.
\end{eqnarray}
The characteristic linear size of the cluster $\ell \sim \langle s \rangle^{1/d}$ thus increases {\em linearly} with time in interface-controlled mechanism.
The corresponding extended volume fraction then obeys a power-law behavior $\phi^e(t) \sim t^{n}$ with the Avrami exponent $n = d+1$. Our results from Fig.~\ref{fig:phi_M}(b) indicates that the early stage of the phase transformation is consistent with the interface-controlled cluster growth, which is also directly verified in Fig.~\ref{fig:size} from our Gutzwiller MD simulations. By taking into account the incubation period, the time dependence of the average cluster size is well approximated by a parabolic function $\overline{s} = a t^2 + b$. In particular, a linear segment with unity slope can be seen in the log-log plot of the characteristic linear size, defined as $\ell = \langle {s} \rangle^{1/2}$, versus the shifted time $t - t^*$, where $t^*$ is the induction time from Fig.~\ref{fig:DT_dep}(b).


The interface-controlled growth mechanism here indicates that electron localization, or local Mott transition to the insulating state, tends to occur at the perimeter of the already formed Mott puddles. Physically, this is because the weakened electron bonding force in the vicinity of the Mott region allows the neighboring atoms to overcome the energy barrier more easily and tunnel into the insulating state through thermal fluctuations. One can also see this mechanism from the local slave-boson Hamiltonian Eq.~(\ref{eq:H_SB}). The kinetic energy, represented by the $\mathcal{M}_\alpha \, \Delta_{i, \alpha}$ term, is reduced in the neighborhood of a Mott cluster due to the reduced $\Delta_{i, \alpha}$ parameter; see Eq.~(\ref{eq:Delta}). The competition with the Hubbard $\mathcal{U}$ thus results in a slave boson state with vanishing~$d_i$. This local competition underlies the microscopic cluster growth process in the Gutzwiller MD simulations.

\begin{figure}
\includegraphics[width=0.96\columnwidth]{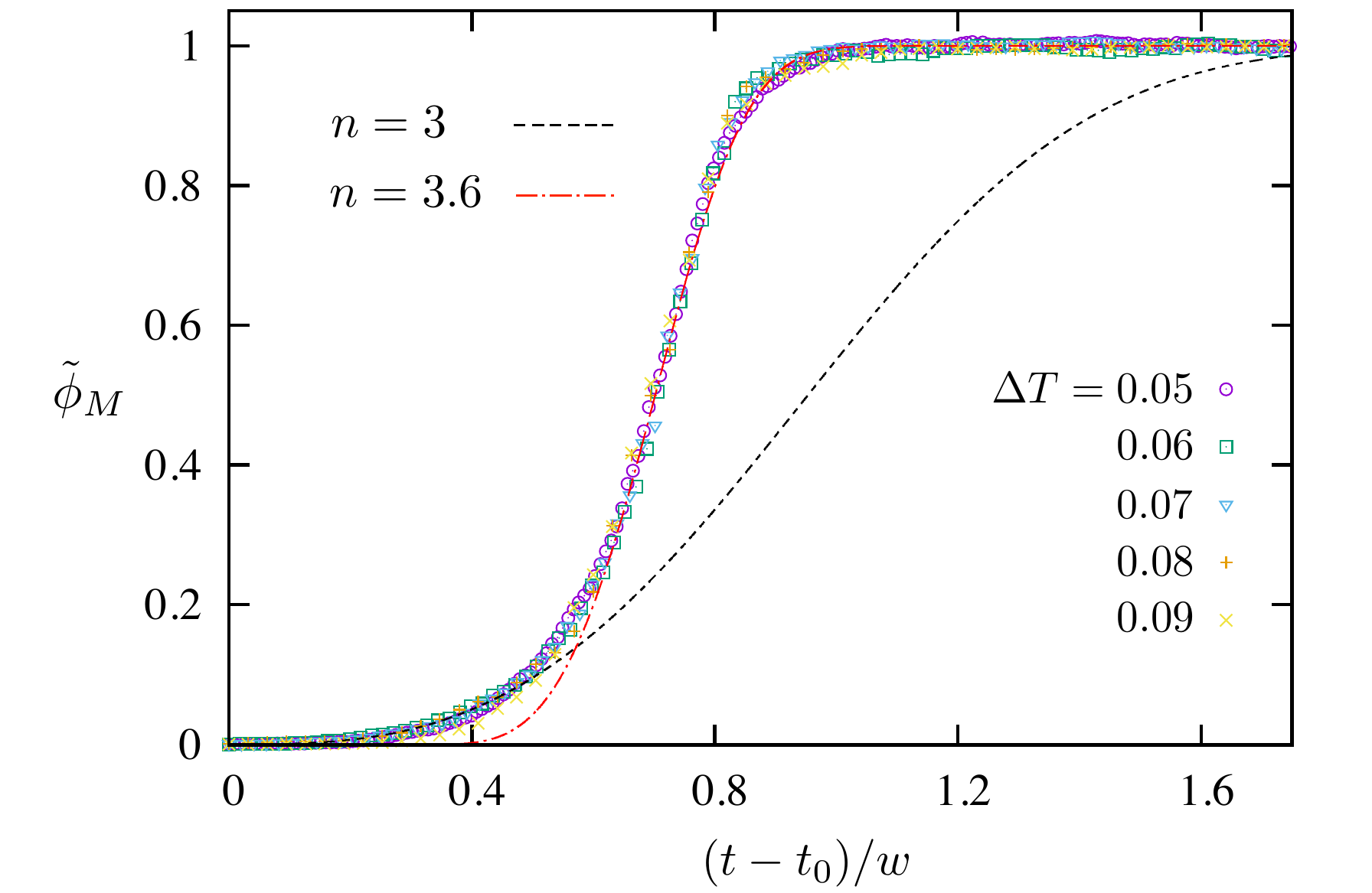}
\caption{(Color online)  
\label{fig:scaled_phi} The Mott volume fraction $\tilde \phi_M$ normalized to its maximum versus the scaled time $(t - t_0)/w$ for varying $\Delta T$. The black dashed line correspond to the Avrami Eq.~(\ref{eq:avrami}) with $n = 3$. The red dash-dotted line is the Avrami equation with $n = 3.6$ shifted in time to fit the large time part of the transformation.
}
\end{figure}

\section{Avalanche behavior}
\label{sec:avalanche}

It is worth noting that the phase transformation of the 2D kinetic Ising model is well described by the standard Avrami equation with $n = 3$~\cite{shneidman99,shneidman99b}. Moreover, cell dynamics simulations based on the stochastic 2D TDGL Eq.~(\ref{eq:TDGL}) with a scalar order parameter gives results quantitatively consistent with the $n = 3$ KJMA model~\cite{iwamatsu08}. In our case, however, the standard KJMA kinetics only lasts a relatively small fraction of the total phase transformation time. After this initial stage of constant nucleation, the transformation speeds up significantly compared with the prediction of the Avrami equation. 
To further quantify the Mott transition kinetics, Fig.~\ref{fig:scaled_phi} shows the normalized Mott volume fraction versus the scaled time $(t - t_0)/w$. Interestingly, by properly choosing the origin $t_0$ and the duration $w$ of the transformation, the numerical data from different $\Delta T$, shown in Fig.~\ref{fig:phi_M}(a), roughly collapse on the same curve, indicating a new universal kinetic behavior for the Mott transition. 
With the origin of the phase transformation fixed at $t = t_0$, there are two parameters, $K$ and $n$, left for fitting the numerical data using Eq.~(\ref{eq:avrami}). As shown in Fig.~\ref{fig:scaled_phi}, the best-fitting result with the Avrami exponent fixed at $n = 3$ only captures the initial part of the transformation curve up to $\tilde{\phi}_M \sim 0.15$. Beyond that, one can see a clear acceleration of the phase transformation compared with the prediction of the KJMA model. In fact, by shifting the origin to a later time, the later stage of the phase transformation seems to be well described by an Avrami equation~(\ref{eq:avrami}) with a larger exponent $n = 3.6$.

\begin{figure}
\includegraphics[width=0.99\columnwidth]{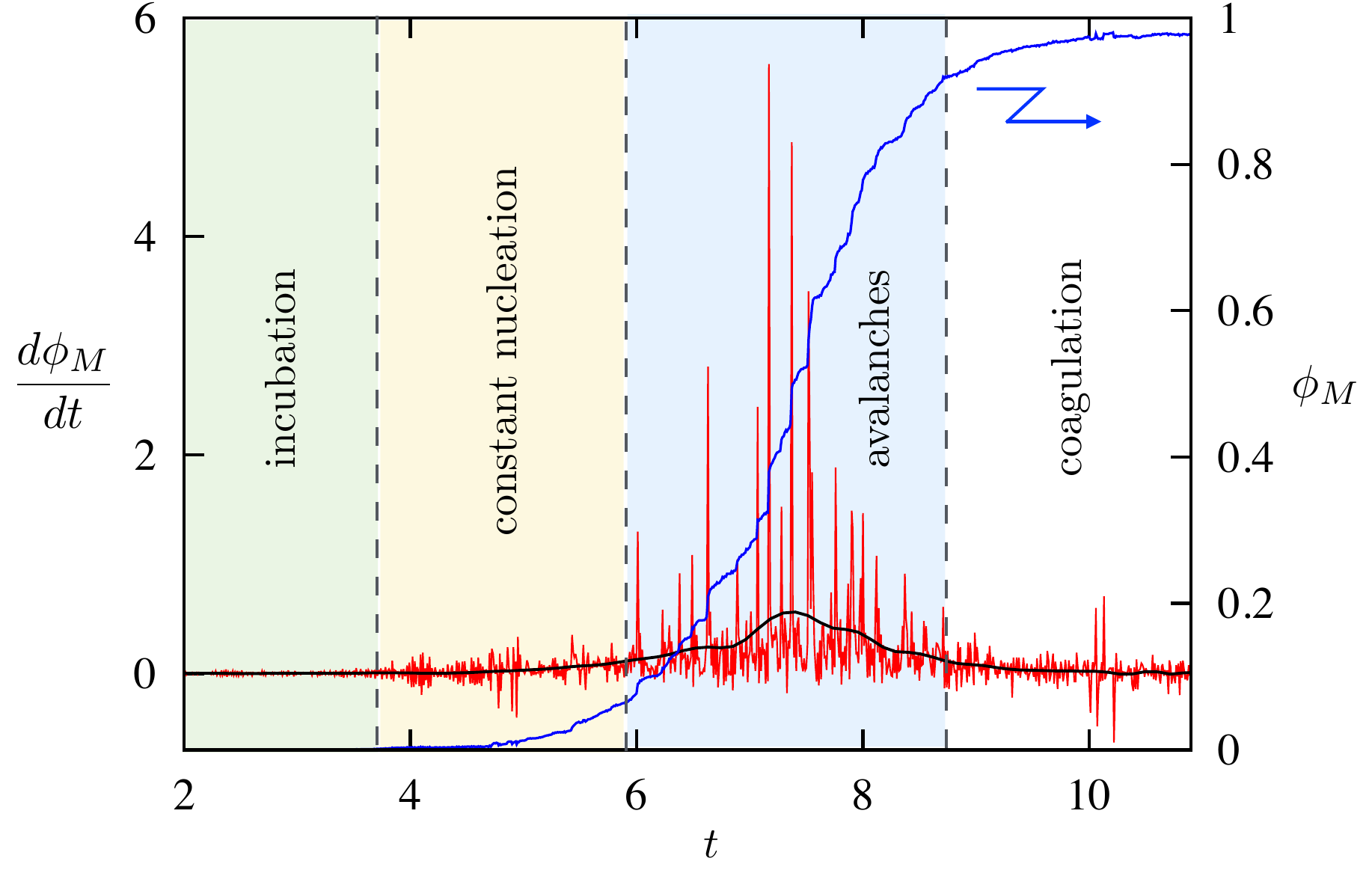}
\caption{(Color online) The transformation rate $d\phi_M/dt$ as a function of time for a particular temperature-quench simulation from $T_i = 0.09$ to $T_f = 0.18$. The smooth black line indicates the averaged transformation rate.  Also shown in the right axis is the time dependence of the Mott volume fraction $\phi_M$. The four distinct stages of the phase transformation are also outlined above.
\label{fig:jumps} 
}
\end{figure}

In addition to the enhanced transformation rate, this new stage of the phase transformation is characterized by kinetics similar to the Barkhausen effect~\cite{durin05,sethna01} in the magnetization or demagnetization processes of ferromagnets. It should be noted that the relatively smooth curves shown in Fig.~\ref{fig:scaled_phi} were obtained by averaging over at least ten independent runs. Individual transformation curves as that shown in Fig.~\ref{fig:jumps} (right axis), however, exhibit step-like features in the accelerated regime of the phase transformation. These steps in $\phi_M$ are also marked by the delta-like peaks in the time dependence of the instantaneous transformation rate $d\phi_M/dt$, see Fig.~\ref{fig:jumps} (left axis). Microscopically, the Barkhausen effect is caused by sudden changes in the size and orientation of ferromagnetic domains~\cite{urbach95,cizeau97,narayan96}. Similarly, the jumps in $\phi_M$ here correspond to large domains of the lattice undergoing simultaneous local Mott transition into the electron localized state. This is also confirmed from the real-space snapshots of the double-occupancy in the vicinity of these steps. Moreover, we find that the formation of these large Mott clusters are not random, but tend to occur in the vicinity of the already existing Mott puddles.

Quenched disorder, which introduces random pinning centers for the propagating domain walls, plays a crucial role in the Barkhausen effect~\cite{sethna01,urbach95,cizeau97,narayan96}. The disorder induces multiple local minima in the configuration space of domain-walls, and the jumps in magnetization correspond to transitions between nearby local minima. Thermal fluctuations, on the other hand, are not necessarily an important factor in the domain-wall dynamics. Remarkably, despite the absence of quenched disorder in our simulations, the transformation to the Mott phase still exhibits the avalanche-like behavior as shown in Fig.~\ref{fig:jumps}. The intermittent bursts of local Mott clusters are likely due to a self-generated disorder in the intermediate mixed-phase states.

This scenario can be qualitatively understood from the viewpoint of the energy landscape of the Mott order parameters. 
At the early stage of the phase transformation, the system moves toward a well defined global minimum in a relatively smooth valley through the nucleation and growth of the Mott clusters. 
As the transformed volume fraction $\phi_M$ increases, more and more lattice sites are disconnected from the rest due to the vanishing effective hopping $t^{\rm eff}_{ij} = \mathcal{R}_i \mathcal{R}_j t_{ij} \to 0$. These diluted bonds in turn introduce off-diagonal disorder to the effective quasi-particle Hamiltonian Eq.~(\ref{eq:H_renorm}) and modify the localization of the electron wavefunctions. After integrating out the electrons, an increasingly rugged energy landscape for the slave-bosons is created by the enhanced self-generated disorder. Similar to the depinning transition of driven domain-walls, the jumps in the Mott volume fraction $\phi_M$ thus corresponds to the transitions between local minima of the resultant energy landscape.




It is worth noting that numerically the avalanche behavior in our MD simulations comes mostly from the Gutzwiller iteration even within a single time-step. The Mott transition of the whole cluster is essentially a collective electronic response. This process is also similar to the avalanche behaviors in, e.g. the random field Ising model (RFIM)~\cite{sethna93,sethna05} or the depinning transition of domain walls~\cite{urbach95,cizeau97,narayan96}. The evolution of these spin systems is often governed by adiabatic or athermal dynamics. The avalanche is initiated by a single spin-flip which triggers a chain-reaction-like response of neighboring spins in one time-step. 

The effective energy functional $\mathcal{E}[\Phi_i]$ of the Mott order parameter can be formally obtained from Eq.~(\ref{eq:free-energy}) by integrating out the electron and the lattice degrees of freedom. Importantly, this process introduces long-range electron-mediated ``interactions" between the local slave-boson variables $\Phi_i$, which could potentially create a complex energy landscape. This nonlocal interaction can be demonstrated by the effective spin model discussed in Appendix~\ref{sec:spin-model}. Assuming particle-hole symmetry, which is a good approximation in our case as the local electron density is very close to half-filling; see Fig.~\ref{fig:hist_db}, one can introduce classical spins $\mathbf S_i$ such that $S_i^x = d_i p_i + p_i d_i$ and $S_i^z = p_i^2 - d_i^2$. The effective Hamiltonian for these spin variables is a random-bond transverse-field Ising model
\begin{eqnarray}
	\mathcal{H}_{\rm eff} = \sum_{\langle ij \rangle} \mathcal{J}_{ij} \,S^x_i S^x_j - H \sum_i S^z_i,
\end{eqnarray}
where the effective longitudinal field $H = U/2$, the nearest-neighbor spin interaction $\mathcal{J}_{ij} = 8 t_{ij} (\rho_{ij} + \rho_{ji})$, and $\rho_{ij}$ is the reduced density matrix of the quasi-particles and depends nonlocally on other spins. In this picture, a significant bond-disorder develops in the mixed-phase states due to inhomogeneous electron wave functions, which in turn results from the inhomogeneous slave-boson or ``spin" configuration.

\begin{figure}
\includegraphics[width=0.97\columnwidth]{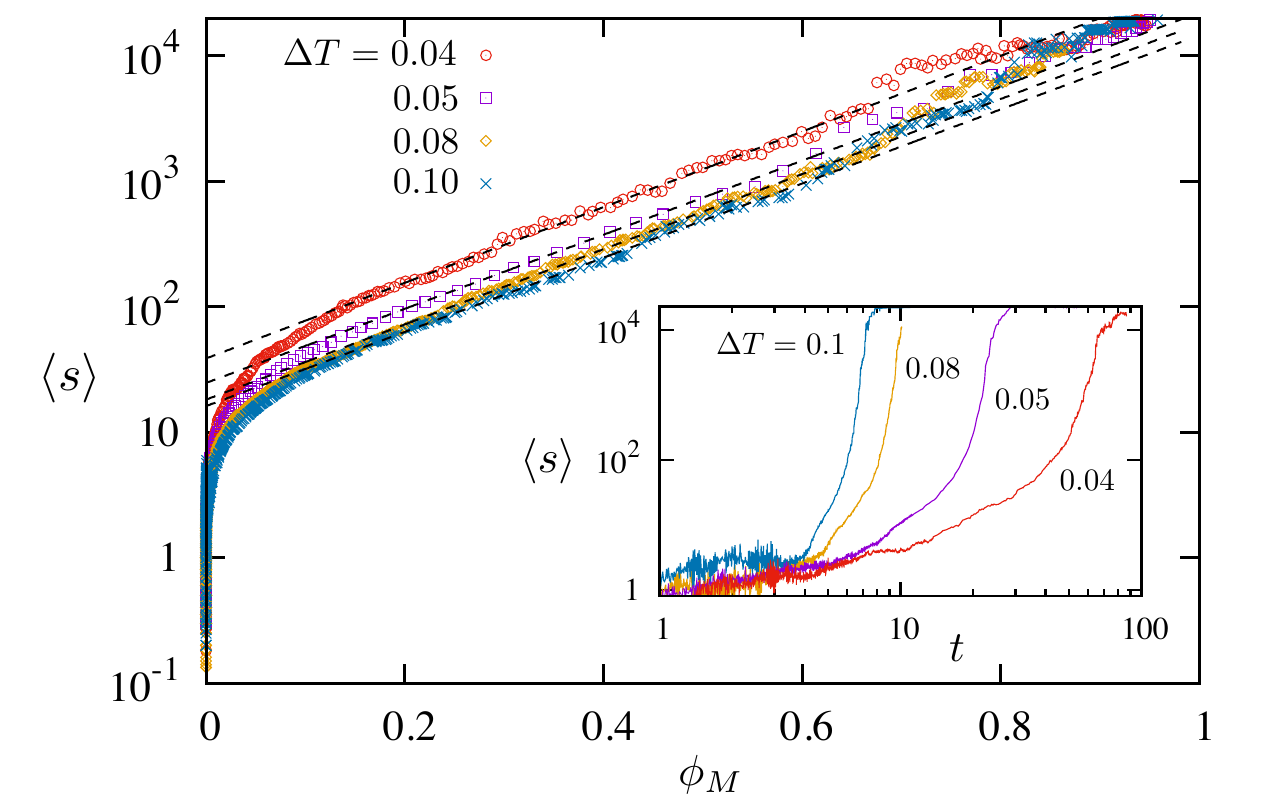}
\caption{(Color online)  
\label{fig:size_phi} The average cluster size $\langle {s} \rangle$ versus the Mott volume fraction $\phi_M$ for varying depth of the temperature quench. The inset shows the log-log plot of $\langle {s} \rangle$ versus time.
}
\end{figure}

The nonlocal nature of the effective interaction also manifests itself in the peculiar growth dynamics of the Mott clusters. We found that the increase of the average cluster size $\langle s \rangle$ is governed neither by a power-law (see inset of Fig.~\ref{fig:size_phi}) or exponential function. Instead, it increases exponentially with the Mott volume fraction
\begin{eqnarray}
	\label{eq:exp_phi}
	\langle s \rangle \sim \exp(\gamma \phi_M),
\end{eqnarray}
where $\gamma \approx 6.8$ -- 6.9 is a constant weakly dependent on~$\Delta T$.
Since the fraction $\phi_M$ describes how many sites have been transformed to the Mott phase in the whole lattice, this dependence indicates that the growth of individual clusters in the avalanche regime depends on the {\em global} property of the system.

Before concluding this section, we outline in Fig.~\ref{fig:jumps} the four distinct kinetic behaviors observed in the Mott transition: (i) the initial incubation period, (ii) the constant nucleation regime, (iii) the accelerated transformation with avalanche behavior, and (iv) the coagulation at the late stage.  In the incubation period, there are no Mott clusters of appreciable sizes and nearly negligible transformation rate.  Once the system enters the quasi-steady state with a constant nucleation rate, the Mott volume fraction steadily increases with a $\phi_M \sim t^3$ dependence, consistent with the interface-controlled cluster-growth mechanism. It is also worth noting that the transition rate $d\phi_M/dt$ exhibits both positive and negative fluctuations, indicating that some Mott clusters would turn back to become metallic throughout the transformation. As the Mott volume fraction reaches $\phi_M \approx 0.15$, the phase transformation kinetics is characterized by avalanche behavior and a non-local cluster-growth phenomenon. And finally, the transition is slowed down at the late stage as the growth of the Mott cluster is blocked by the coagulation effect.


\section{Structure of the mixed-phase states}
\label{sec:structure}

\begin{figure}
\includegraphics[width=0.99\columnwidth]{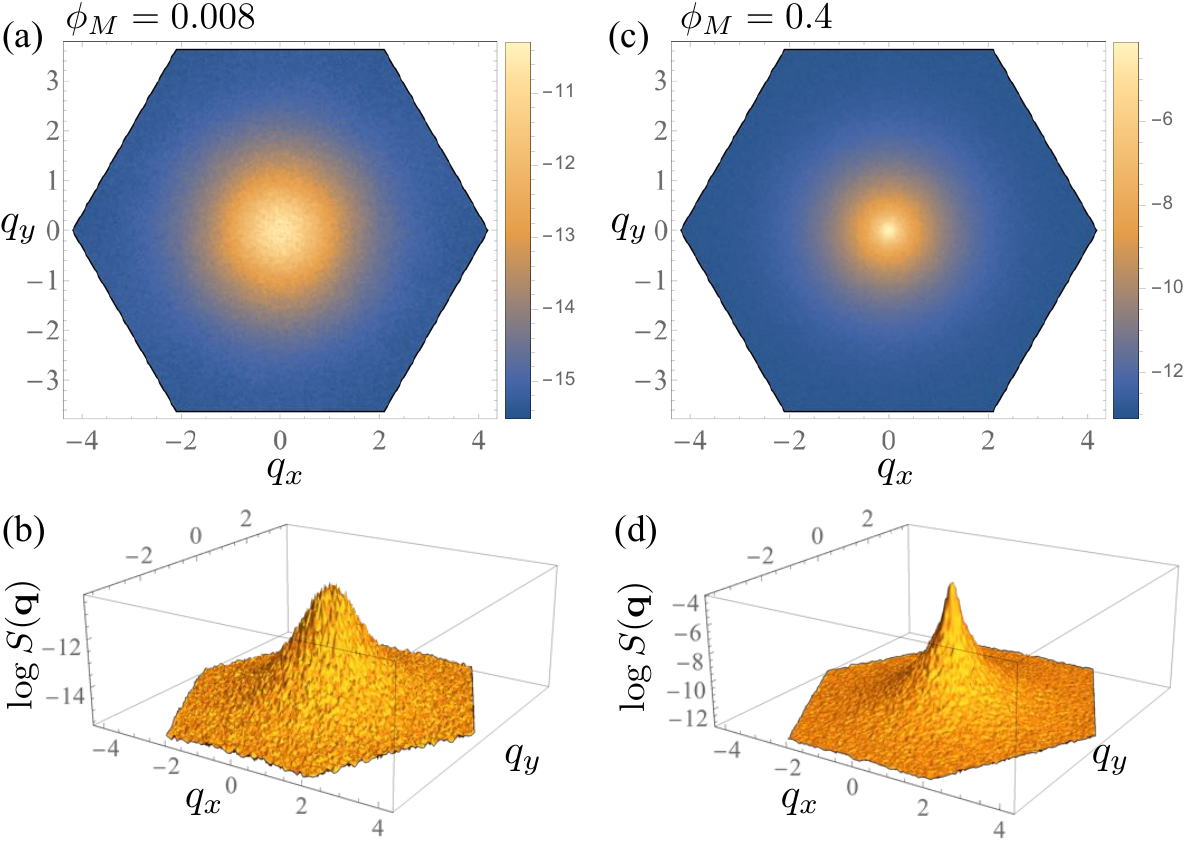}
\caption{(Color online)  
\label{fig:Sq} The logarithm of the structure factor~$S(\mathbf q)$ at (a), (b) $\phi_M = 0.008$  and (c), (d) $\phi_M = 0.4$ of the temperature quench shown in Fig.~\ref{fig:trace}. They correspond to simulation time $t = 15$  $t = 52$, respectively.
}
\end{figure}

We next characterize the structure of the intermediate mixed-phase states during the phase transformation. In order to compare with results from the kinetic Ising model, we first map the electronic state to an effective Ising configuration by introducing a binary variable $\sigma_i$ such that $\sigma_i = +1$ if the corresponding site-$i$ is metallic, and $\sigma_i = -1$ if it is a Mott insulator. Again, numerically we define a site to be a Mott insulator if its double-occupancy $D_i < 10^{-8}$. Our central interest here is the time-dependent two-point correlation function 
\begin{eqnarray}
	G(\mathbf r_{ij}, t) = \langle \delta\sigma_i(t) \delta\sigma_j(t) \rangle = \langle \sigma_i(t) \sigma_j(t) \rangle - \langle \sigma(t) \rangle^2,
\end{eqnarray}
where $\delta\sigma_i \equiv \sigma_i - \langle \sigma \rangle$ denotes the fluctuations relative to the ``ferromagnetic'' order. In this mapping, the metallic and Mott phases correspond to the two opposite polarized Ising-states with $\langle \sigma \rangle > 0$ and $\langle \sigma \rangle < 0$, respectively. The Fourier transform of the correlation function gives the structure factor
\begin{eqnarray}
	S(\mathbf q, t) = \sum_{\mathbf r} G(\mathbf r, t) e^{i \mathbf q \cdot \mathbf r} = \Bigl \langle \Bigl | \sum_i \delta \sigma_i(t) \,e^{i \mathbf q\cdot \mathbf r_i} \Bigr|^2 \Bigr\rangle,
\end{eqnarray}
During the phase transformation, the presence of a finite ferromagnetic order introduces a huge peak at $\mathbf q = 0$. In order to see details of the structure factor, we plot in Fig.~\ref{fig:Sq} the logarithm of $S(\mathbf q)$ computed at two different simulation times of the temperature quench shown in Fig.~\ref{fig:trace}. An emergent circular symmetry of the structure factor indicates an isotropic phase-separated states. The central peak becomes sharper and stronger as the phase transformation progresses.

\begin{figure}
\includegraphics[width=0.9\columnwidth]{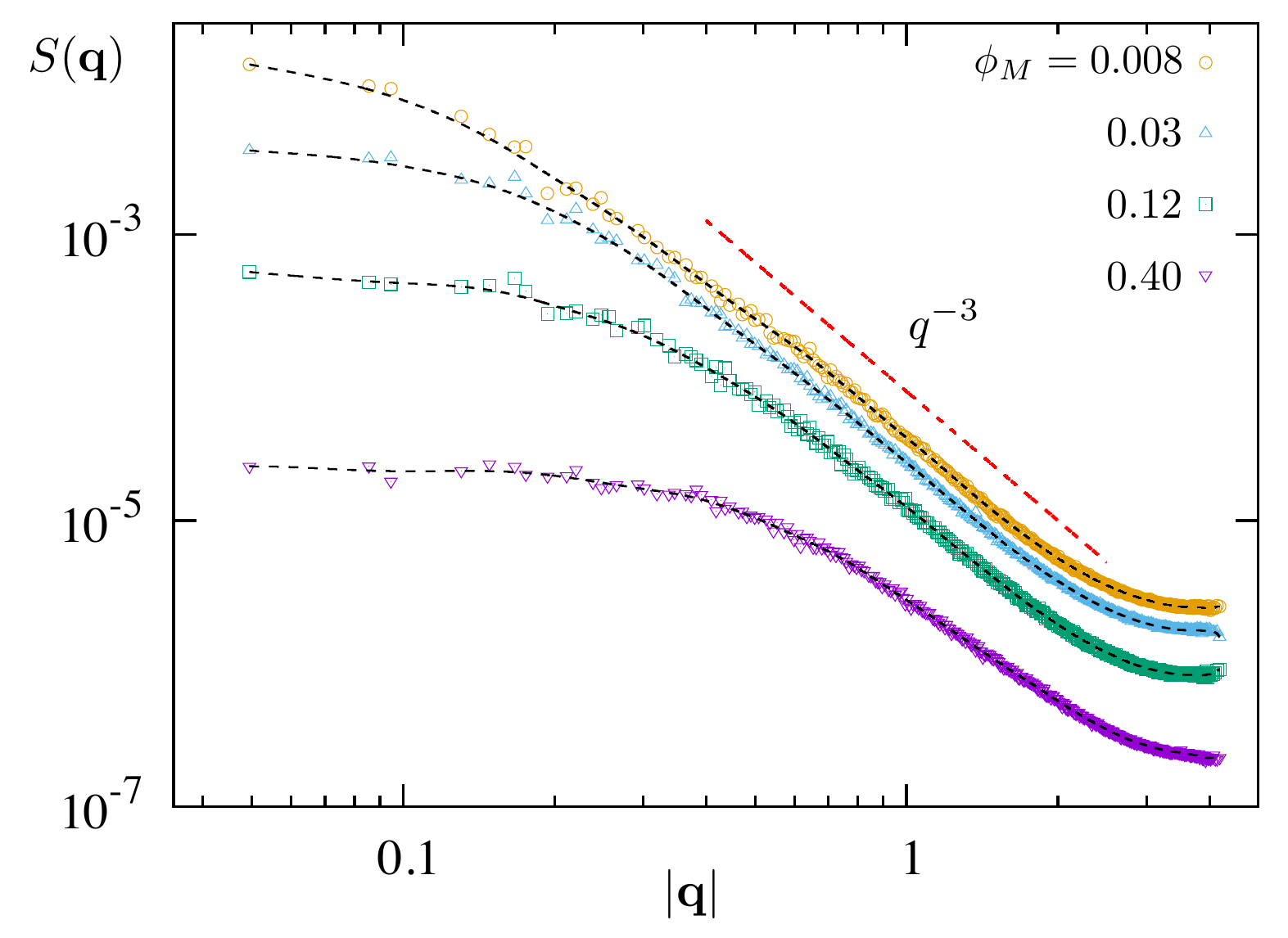}
\caption{(Color online)  
\label{fig:sq_cmp} Log-log plot of the circularly averaged structure factor~$S(\mathbf q)$ versus the $|\mathbf q|$ at four different simulation times. The black dashed lines are guide for the line. The $1/q^3$ power-law dependence is shown by the red long dashed line.
}
\end{figure}

\begin{figure}
\includegraphics[width=0.99\columnwidth]{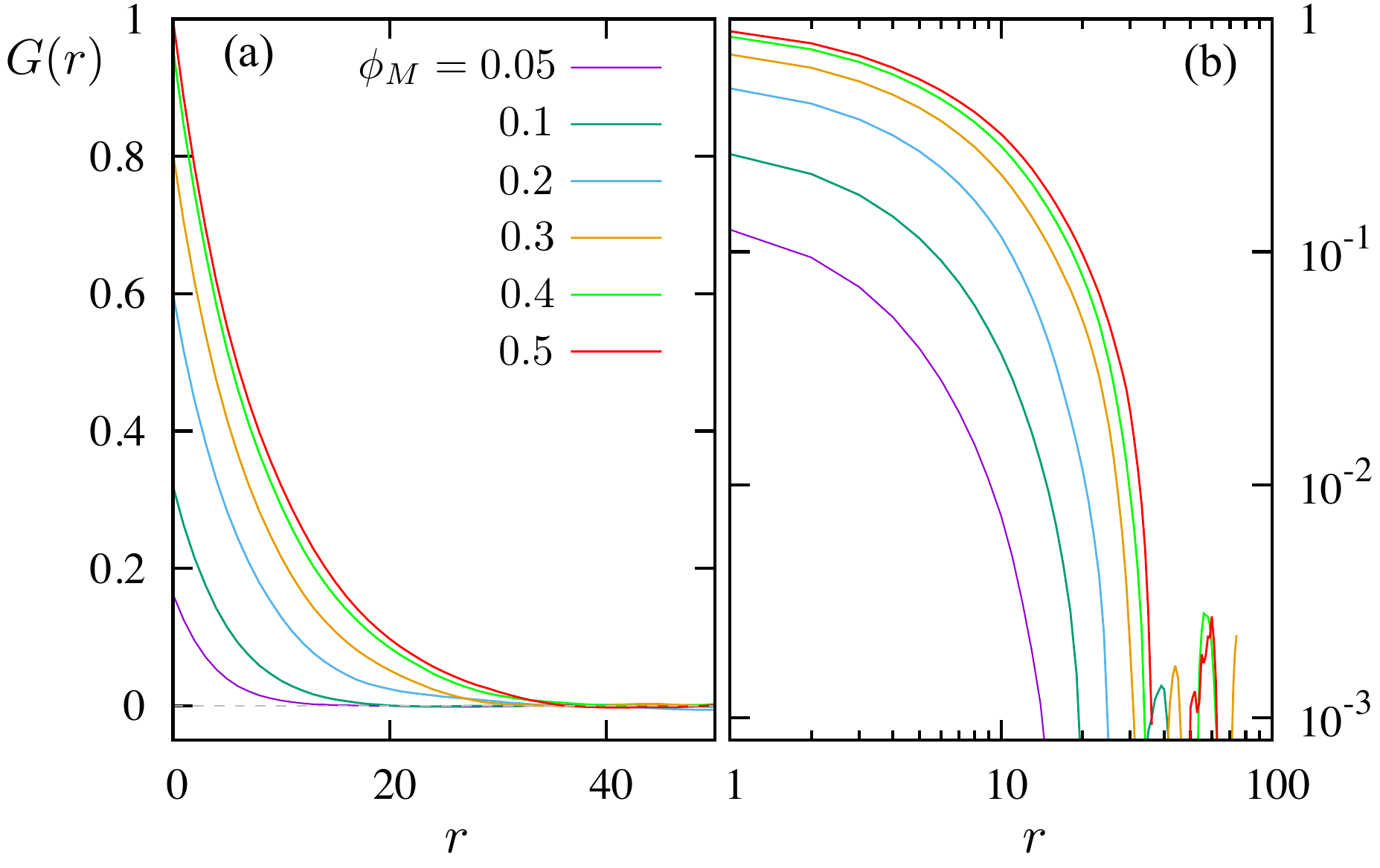}
\caption{(Color online)  
\label{fig:gr} (a) Correlation function $g(r)$ at varying Mott volume fraction $\phi_M$ from the temperature quench simulation shown in Fig.~\ref{fig:trace}. (b) The same curves shown in log-log plot.
}
\end{figure}

Fig.~\ref{fig:sq_cmp} shows the circularly averaged structure factor as a function of the modulus of $\mathbf q$ at four different Mott volume fractions. Although not much information can be obtained from the small $\mathbf q$ behaviors, the large wavevector dependence is consistent with the Porod's law which states that the structure factor for a two-phase system with interfaces of negligible thickness should behave as $S(q) \sim 1/q^{d+1}$~\cite{porod82}, where $d$ is the spatial dimension. The deviation from the Porod's law at the Brillouin zone boundary results from thermal fluctuations and finite width of the metal-insulator interfaces. 
The power-law $q$-dependence also indicates a short-distance singularity of the correlation function $G(\mathbf r) \approx (1 - \langle \sigma \rangle^2)\,(1 - A | \mathbf r|)$ at small $|\mathbf r|$, where $A$ is a constant. This linear-$r$ dependence at short distances is also confirmed in our simulations; see Fig.~\ref{fig:gr}(a) for~$G(r)$ computed at various Mott volume fractions~$\phi_M < 0.5$. 

The correlation length $\xi$ estimated from these correlation curves  is found to increase monotonically with time before reaching the mid-point $\phi_M = 0.5$ of the phase transformation. A more systematic determination of $\xi$ will be discussed below. Importantly, despite the emergence of a transient power-law size distribution of the Mott clusters discussed in Sec.~\ref{sec:nucleation}, the correlation length remains finite throughout the phase transformation, as evidenced by the log-log plot of the correlation functions shown in Fig.~\ref{fig:gr}(b). This result thus means that the probability of two separate sites being on the same large cluster is exceedingly small. A similar example is the 3D Ising model, in which a power-law distribution of geometrical spin clusters occurs at a temperature which is different from the critical point.

\begin{figure}
\includegraphics[width=0.99\columnwidth]{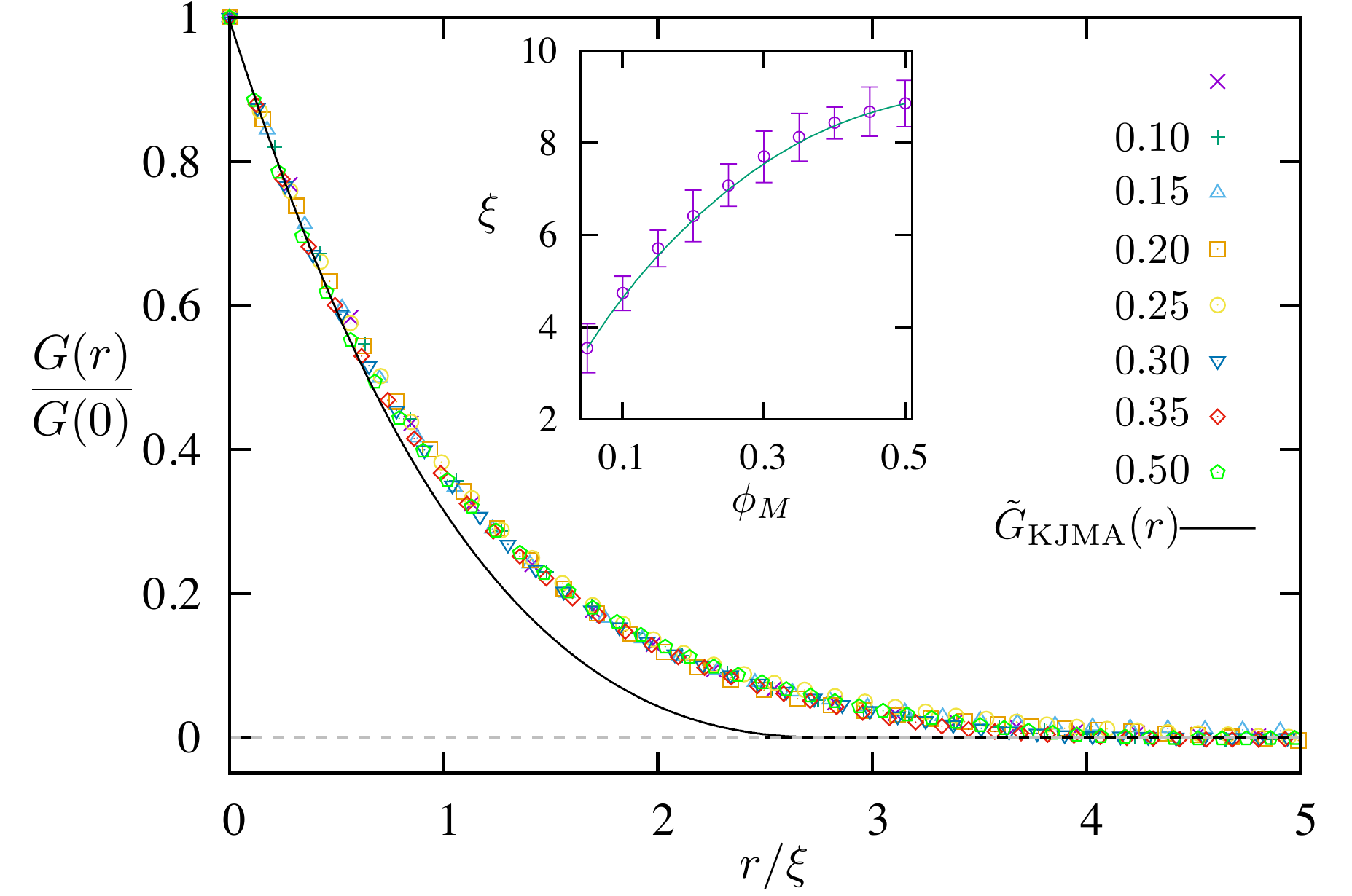}
\caption{(Color online)  
\label{fig:gr_scaled} Data collapsing from the normalized $g(r)$ versus rescaled distance $r/\xi$. Also shown is the analytical correlation function based on the KJMA model $\tilde{G}_{\rm KJMA}(r) = \exp[C \Psi_2(r/\zeta)] - 1$. The inset shows the length scaling factor $\xi$ (in arbitrary units) versus the Mott volume fraction. 
}
\end{figure}

On the other hand, the correlation curves at different times seem to exhibit a self-similarity. To investigate potential universal behaviors, we compute the normalized correlation functions $\tilde G(r, t) \equiv G(r, t)/G(0, t)$ and plot them against a rescaled distance $r \to r/\xi$ in Fig.~\ref{fig:gr_scaled}. The rather nice data points collapsing in the plot strongly suggests a universal correlation function for the mixed-phase state of the Mott transition. The scaling factor $\xi$, determined from the condition $\tilde G(\xi) = 1/e$, also provides a measure of the finite correlation length. The inset of Fig.~\ref{fig:gr_scaled} shows the numerical $\xi$ versus the Mott volume fraction. Interestingly, the correlation length $\xi$ is at most of the order of 10 lattice constants.

Fig.~\ref{fig:gr_scaled} also shows the comparison of the numerical universal correlation function with the analytical formula obtained by Sekimoto based on the interface-controlled mechanism within the KJMA theory~\cite{sekimoto84,sekimoto86}. As discussed in Sec.~\ref{sec:interface}, the radius of a spherically shaped nuclei in $d$-dimensions increases linearly with time, i.e. $R(t) = R_0 + v t$, where $v$ is the linear growth rate. Assuming negligible initial size $R_0$, the finite growth velocity implies that $G(r, t) = 0$ for $r > 2 v t$. For separation of two points within the diameter $2 R(t)$ of a nucleating cluster, the correlation function is finite and its value depends on the random distribution of the critical nuclei. Detailed calculation gives~\cite{sekimoto86} 
\begin{eqnarray}
	\label{eq:gr}
	G(r, t) =4  [1- \phi(t)]^2  \!\left\{\exp\!\left[\mathcal{I}\, v^d t^{d+1} \Psi_d\!\left(\frac{r}{2vt}\right)\right] - 1\right\}, \,\,
\end{eqnarray}
for $r < 2 vt$. Here $\phi(t)$ is the transformed volume fraction in Eq.~(\ref{eq:avrami}), and the function $\Psi_d(x)$ in 2D is
\begin{eqnarray}
	\Psi_2(x) \!=\! \frac{2}{3} \!\left[ \cos^{-1}\! x - 2 x \sqrt{1-x^2} + x^3 \ln\! \left(\frac{1 + \sqrt{1-x^2}}{x} \right) \right]\!. \nonumber
\end{eqnarray}
It is worth noting that this analytical expression agrees very well with the kinetic Monte Carlo simulation results of the phase transformation in dynamical Ising model~\cite{ramos99}. The normalized correlation function can be characterized by two parameters: $\tilde G(r) = \exp[C \, \Psi_2(r/\zeta)] - 1$. Here $C$ and $\zeta$ are determined from the conditions $\tilde G(0) = 1$ and that the gradient $d\tilde G/dr|_0$ equals the slope of the linear segment of the numerical correlation function at small $r$. As shown in Fig.~\ref{fig:gr}(b), although the short-range correlation of the mixed-phase states can be satisfactorily described by the KJMA model, the correlation function is appreciably enhanced at larger distances. The discrepancy from the analytical formula~(\ref{eq:gr}) again indicates the break down of interface-controlled mechanism with constant domain growth rate, and the enhanced correlation can be attributed to the avalanche of Mott clusters discussed in Sec.~\ref{sec:avalanche}. 


\section{Conclusion and outlook}
\label{sec:outlook}

To summarize, as a first step toward understanding the intricate dynamical textures during Mott metal-insulator transition of correlated electron materials~\cite{qazilbash07,liu13,madan15,lupi10,qzailbash11,mcleod16,stinson18,mattoni16,ronchi18,preziosi18,singer18}, we have performed systematic and comprehensive large-scale dynamical simulation of thermally driven Mott transition in a Hubbard-type model. Our multi-scale approach to this difficult problem is based on an efficient integration of the linear-scaling kernel polynomial method (KPM) and a novel Gutzwiller/slave-boson molecular dynamics (MD) method. Contrary to  phenomenological approaches for inhomogeneity in metal-insulator transition, such as the kinetic Ising model or time-dependent Ginzburg-Landau equation, our Gutzwiller MD simulation offers, for the first time, microscopic details of the complex phase transformation dynamics. In our multi-scale modeling, the slave-boson amplitudes, which represent collective electron behaviors such as the double-occupancy, serve as the order parameters for the Mott transition. To include the crucial electron correlation effects, the electron degrees of freedom are integrated out {\em on the fly} by solving the Hubbard model with efficient Gutzwiller solver at every MD time-step. The Langevin dynamics is used to model lattice fluctuations, which give rise to the relaxation dynamics and the stochastic force for the order-parameter fields. 

Our MD simulations of temperature quench process reveal that the metal-to-insulator phase transformation is initiated by the nucleation of Mott droplets in a metallic background. The initial fluctuations of the Mott clusters are well described by the classical droplet model~\cite{gunton83}. After an incubation period, whose duration decreases with increasing depth of the temperature quench, the system enters a nonequilibrium quasi-steady state with a constant nucleation rate, consistent with the picture of the classical Becker-D\"oring-Zeldovich nucleation theory~\cite{becker35,zeldovich43}. The total number of the nuclei as well as their characteristic linear size increase linearly with time in this constant nucleation regime. This in turns points to an underlying interface-controlled cluster-growth mechanism, i.e. the growth rate of the nuclei is proportional to the surface area, which is also directly confirmed in our MD simulations.

The time evolution of the transformed Mott volume fraction $\phi_M$ follows a universal function of the scaled time, after taking into account the initial incubation period. Interestingly, the classical theory of phase transition kinetics developed by Kolmogorov, Johnson, Mehl, and Avrami (KJMA model)~\cite{kolmogorov37,avrami39,avrami40,johnson39} only describes the early constant-nucleation stage of this universal kinetics of the Mott transition. As the Mott volume fraction exceeds $\phi_M^* \approx 0.15$, the transformation speed picks up compared with the KJMA model. This accelerated transformation stage is also characterized by intermittent bursts of large-size Mott clusters and a nonlocal nature of the cluster growth. We argue that this avalanche behavior, similar to the Barkhausen effect, could be attributed to self-generated disorder for the order-parameter field. From this new perspective, it would be interesting to investigate the effects of  emergent disorder  on the quasi-particle wave function in the mixed-phase states. Such study will help understanding the mechanism of the unusual transition dynamics. Much insight can be drawn from the extensive research on the interplay between Anderson localization and Mott transition~\cite{dobrosavljevic97,dobrosavljevic03,byczuk05,aguiar09,tanaskovic03,andrade09,andrade10}. We hope future studies will shed light on the peculiar relation Eq.~(\ref{eq:exp_phi}) that describes an exponential dependence of the average cluster size on the Mott volume fraction in this dynamical regime.

Finally, we have also characterize the structure of the transient mixed-phase states. The structure factor shows the telltale Porod's power-law at large wavevectors, indicating sharp metal-insulator interfaces in the phase-separated states. A universal two-point correlation function is obtained after properly scaling the distance with respect to the correlation length. Although a power-law cluster-size distribution emerges at roughly the midpoint of the phase transition, the correlation length remains finite through out the transformation. 

A detailed study of the the metal-insulator coexistence regime in VO$_2$ was recently carried out by applying the critical cluster techniques to the experimental near-field images~\cite{liu16}. Interestingly, a robust power-law cluster-size distribution $\mathcal{D}(s) \sim 1/s^\tau$ was observed over a wide temperature range with an exponent $\tau = 1.92 \pm 0.11$~\cite{liu16}. Other geometrical characterizations such as the volume fractal dimension and hull fractal dimension were also computed. After comparing these exponents with that of potential universality classes, the authors of Ref.~\cite{liu16} concluded that the MIT in VO$_2$ is consistent with the system passing near criticality of the random field Ising model (RFIM)~\cite{sethna93,sethna05} as temperature is varied. Although the criticality of the Mott transition is expected to be the Ising universality class in the absence of quenched disorder, as in our case, the results of our cluster analysis in Sec.~\ref{sec:nucleation} seem to be consistent with both scenarios. We plan to clarify this issue  and to study the crossover phenomena between the thermal and disorder-dominated Mott transitions by carrying out similar Gutzwiller MD simulations on Hubbard models with quenched randomness in the future.

Another recent nano-imaging experiment on V$_2$O$_3$ offers a first dynamical picture of a photo-induced Mott insulator-to-metal transition in real materials~\cite{ronchi18}. The volume fraction of the transformed metallic phase can be estimated from the reflectivity data. Interestingly, a two-stage transformation was observed from the time evolution of the metallic volume fraction. The initial stage is well described by the KJMA model with an exponent $n = 2$, suggesting a 2D domain growth with pre-existing nuclei which are quickly exhausted~\cite{christian75}. Contrary to our case, the transformation after this initial stage is slowed down. The reduced Avrami exponent, ranging from 0 to $n \approx 3/4$ depending on the temperature, was suggested to result from a martenistic transition characterized by highly anisotropic domain growth~\cite{cahn56}. To shed light on these intriguing results, we plan to conduct a systematic study on the kinetics of Mott transition in the reverse direction, i.e. the transformation from the Mott insulator to the correlated metal. Another intriguing result is the striped domains in the phase separated states of V$_2$O$_3$~\cite{mcleod16,ronchi18}. Similar stripe patterns have also been observed in the mixed-phase states of VO$_2$~\cite{liu13}. It is likely that structural transition or dimerization, which are not included in our model, plays an important role in the stripe pattern and the martenistic transformation dynamics.


As discussed in the Introduction, the inclusion of lattice degrees of freedom through molecular dynamics in our modeling is mainly to introduce relaxational dynamics as well as stochastic forces to the slave-boson order parameter field. In this regard, the Hubbard-Peierls model is a simple system that serves the purpose. However, this model is over-simplified as a realistic description of lattice dynamics and structural transition in many Mott materials.  In future study, we plan to apply the Gutzwiller MD to Hubbard-type models that explicitly include structural distortions. One simple model of this sort is the dimer-Hubbard model, in which the structural phase transition is represented by the dimerization. Another future direction is to incorporate the Hubbard-type interactions into the so-called tight-binding MD method (TBMD)~\cite{khan89,wang89,goedecker94,koskinen09}. In this semi-empirical formalism, the effective tight-binding Hamiltonian is determined from the instantaneous atomic configuration through a hopping parameter $t(r_{ij})$ and a classical two-body potential $\phi(r_{ij})$, both dependent on the interatomic distance. For example, an $s$-orbital TBMD was employed in our previous work~\cite{chern17} to construct a liquid Hubbard model. However, stabilization of a crystal phase requires anisotropic hopping with higher-order orbitals. Such TBMD-Hubbard approach without the restriction to an underlying fixed lattice also make it possible to investigate the effects and dynamics of structural defects in a Mott transition.

Finally, we remark that the limitations of the Gutzwiller and the slave-boson methods are well known. Notably among them are the neglect of quantum fluctuations and its inability to describe the incoherent electronic excitations. Moreover, while the spatial fluctuations and inhomogeneity are included in the real-space Gutzwiller method, as used in our simulations, the inter-site correlations are ignored. This drawback is shared by the single-site DMFT. While multi-site or cluster Gutzwiller has been proposed~\cite{ayral17,lee19}, a self-consistent treatment is required for application to inhomogeneous electronic states. On the other hand, the Gutzwiller approach as a mean-field theory of the Mott transition offers a natural framework for multi-scale modeling of the correlated electron system. Also importantly, the GA is probably one of the few many-body techniques that can be feasibly combined with large-scale MD simulations. To go beyond the GA, the real-space DMFT~\cite{potthoff99,freericks04,helmes08} as well as variational Monte Carlo (VMC) can be potentially used as the many-body solver of the quantum MD simulations. In particular, VMC has been used to study the Mott transition in disordered Hubbard model~\cite{pezzoli09,pezzoli10}. There have also been remarkable developments in the VMC-MD method and its applications in recent years~\cite{magro96,delaney06,attaccalite08,mazzola12}. One interesting direction for the future study would be to apply the VMC-MD to the same model Hamiltonian considered in this work. The time-consuming many-body calculations required in these sophisticated methods naturally limit the lattice sizes that can be feasibly simulated. As demonstrated in our recent works~\cite{ma19,suwa19}, combining the advanced many-body techniques, such as real-space DMFT or VMC, with the machine learning methods offers a promising solution for large-scale dynamical simulations of correlated electron systems.

\bigskip
\begin{acknowledgments}
The author thanks C. D. Batista, K. Barros, and G.~Kotliar for fruitful discussions and collaborations on related projects. The author also thanks K. Barros for the help with the kernel polynomial method, and N. Lanata and V. Dobrosavljevi\'c for insightful discussions on the real-space Gutzwiller/slave-boson calculation. This work is partially supported by the Center for Materials Theory as a part of the Computational Materials Science (CMS) program, funded by the US Department of Energy, Office of Science, Basic Energy Sciences, Materials Sciences and Engineering Division. The authors also acknowledge Advanced Research Computing Services at the University of Virginia for providing technical support that has contributed to the results in this paper.
\end{acknowledgments}

\bigskip

\appendix

\widetext

\section{Gutzwiller/Slave Boson method}
\label{sec:GA}

Here we outline details of the Gutzwiller solver used in the MD simulations. In terms of the slave boson coherent state $\Phi_i = (e_i, p_{i\uparrow}, p_{i\downarrow}, d_i)$ and the quasi-particle density matrix $\rho^{\rm qp}$, the free energy in Eq.~(\ref{eq:free-energy}) becomes
\begin{eqnarray}
	\mathcal{F}[\rho^{\rm qp}, \Phi] &=& \sum_{\langle ij \rangle}\sum_\alpha \left(t_{ij}  \mathcal{R}^{\;}_{i\alpha} \mathcal{R}^*_{j\alpha}\, \rho_{j\alpha,i\alpha} 
	+ t_{ji} \mathcal{R}^*_{i\alpha} \mathcal{R}^{\;}_{j\alpha} \, \rho_{i\alpha,j\alpha} \right) + U \sum_i D_i  
	 + \,\, \sum_{i, \alpha} \mu_{i\alpha} \left(n_{i,\alpha} - \rho_{i\alpha, i\alpha} \right) \nonumber \\
	& & + T \, {\rm Tr}(\rho^{\rm qp} \ln \rho^{\rm qp}) 
	 + \,\, T \sum_i \left[ |e_i|^2 \ln\frac{|e_i|^2}{\Pi_{i,e}} + |p_{i\uparrow}|^2 \ln \frac{|p_{i \uparrow}|^2}{\Pi_{i,\uparrow}} + + |p_{i,\downarrow}|^2 \ln \frac{|p_{i \downarrow}|^2}{\Pi_{i\downarrow}}
	+ |d_i|^2 \ln \frac{|d_i|^2}{\Pi_{i, d}} \right].
\end{eqnarray}
Here we have defined the two-point correlation function, $\rho_{i\alpha, j\beta} \equiv \langle c^\dagger_{j\beta} c^{\;}_{i\alpha} \rangle = {\rm Tr}(\rho^{\rm qp}\, c^\dagger_{j\beta} c^{\;}_{i\alpha} )$, which is also called the single-particle reduced density matrix. The particle number $n_{i\alpha}$, the local double occupancy, And the renormalization factors are given by [cf. Eq.~(\ref{eq:R})]
\begin{eqnarray}
	\label{eq:R_2}
	n_{i, \alpha} = |p_{i, \alpha}|^2 + |d_i|^2, \qquad D_i = |d_i|^2, \qquad 
	\mathcal{R}_{i,\alpha} = \frac{ p^*_{i, \alpha} e^{\;}_i + d_i^* p^{\;}_{i, \overline{\alpha}} }{\sqrt{n_{i\alpha} (1 - n_{i\alpha})}}.
\end{eqnarray}
The symbol $\overline{\alpha}$ denotes the time-reversed spin direction of $\alpha$. The uncorrelated local distribution $\Pi^0_i$ can be expressed in terms of the local electron density
\begin{eqnarray}
	\label{eq:P0}
	\Pi_{i, e} = (1-n_{i, \uparrow})(1-n_{i, \downarrow}), \quad \Pi_{i, \uparrow} = n_{i, \uparrow} (1- n_{i, \downarrow}), \quad 
	\Pi_{i, \downarrow} = n_{i, \downarrow} (1- n_{i, \uparrow}), \quad \Pi_{i, d} = n_{i,\uparrow} n_{i, \downarrow}.
\end{eqnarray}
In particular, both $\mathcal{R}_{i\alpha}$ and $\Pi_i$ should be viewed as functions of the slave boson amplitudes through $n_{i\alpha}$. Minimization of the $\mathcal{F}$ with respect to the quasi-particle density matrix amounts to solving the eigenvalue problem $H^{(1)} \varphi^{(m)} = \varepsilon_m \varphi^{(m)}$, where $H^{(1)}$ is the ``first-quantized" matrix of the renormalized quasi-particle Hamiltonian given in Eq.~(\ref{eq:H_renorm}), i.e. $\mathcal{H}^{\rm qp} = \sum_{i\alpha, j\beta} c^\dagger_{i\alpha} H^{(1)}_{i\alpha, j\beta} c^{\;}_{j\beta}$. The reduced density matrix can be expressed as $\rho = f_{\rm FD}(H^{(1)} - \mu)$, or explicitly in terms of the real-space eigenstates:
\begin{eqnarray}
	\rho_{i\alpha, j\beta} = \sum_{m} f_{\rm FD}(\varepsilon_m - \mu) \varphi^{(m)*}_{i\alpha}\varphi^{(m)}_{j\beta},  
\end{eqnarray}
where $f_{\rm FD}(\varepsilon)$ is the Fermi-Dirac function. However, calculation of $\rho$ based on exact diagonalization is infeasible for large-scale simulations. Instead, here we adopt the linear-scaling kernel polynomial method (KPM)~\cite{barros13,wang18} to solve the quasi-particle density matrix. 

The minimization with respect to the variational variables can be achieved by solving the coupled equations $\partial \mathcal{F} / \partial \Phi^\dagger_i = 0$ subject to the normalization condition $ |e_i|^2 + |p_{i\uparrow}|^2 + |p_{i \downarrow}|^2 + |d_i|^2 = 1$. Explicit calculations lead to the following expressions
\begin{eqnarray}
	\label{eq:dFde}
	\frac{\partial \mathcal{F}}{\partial e^*_i} = \sum_{\beta = \uparrow, \downarrow}  \Delta^*_{i,\beta} \frac{\partial \mathcal{R}^*_{i,\beta}}{\partial e^*_i} 
	+ T \left( \ln\frac{|e_i|^2}{\Pi_{i, e}} + 1\right) e_i = \sum_{\beta = \uparrow, \downarrow} \frac{\Delta^*_{i,\beta}\, p_{i,\beta}}{\sqrt{n_{i\beta} (1 - n_{i\beta})}} + T \left( \ln\frac{|e_i|^2}{\Pi_{i, e}} + 1\right) e_i, 
\end{eqnarray}
\begin{eqnarray}
	\label{eq:dFdp}
	\frac{\partial \mathcal{F}}{\partial p^*_{i\alpha}} &=&  \Delta_{i,\alpha} \frac{\partial \mathcal{R}_{i,\alpha}}{\partial p^*_{i\alpha}}
	+ \Delta^*_{i, \overline{\alpha}} \frac{\partial \mathcal{R}^*_{i, \overline{\alpha}}}{\partial p^*_{i\alpha}}
	+ \left( \Delta_{i,\alpha} \frac{\partial \mathcal{R}_{i,\alpha}}{\partial n_{i\alpha}} + \Delta^*_{i, \alpha} \frac{\partial \mathcal{R}^*_{i, \alpha}}{\partial n_{i\alpha}} \right) \frac{\partial n_{i\alpha}}{\partial p^*_{i \alpha}} 
	+ \mu_{i, \alpha} p_{i\alpha}
	+ T \left( \ln\frac{|p_{i \alpha}|^2}{\Pi_{i, \alpha}} + 1\right) p_{i\alpha} \nonumber \\
	& & - T \left( |e_i|^2 \frac{\partial \ln \Pi_{i, e}}{\partial n_{i,\alpha}} 
	+  |p_{i\uparrow}|^2 \frac{\partial \ln \Pi_{i, \uparrow}}{\partial n_{i,\alpha}} 
	+  |p_{i\downarrow}|^2 \frac{\partial \ln \Pi_{i, \downarrow}}{\partial n_{i,\alpha}} 
	+  |d_i|^2 \frac{\partial \ln \Pi_{i, d}}{\partial n_{i,\alpha}}  \right) \frac{\partial n_{i,\alpha}}{\partial p^*_{i \alpha}} \nonumber \\
	& = & \frac{\Delta_{i, \alpha} \, e_i}{\sqrt{n_{i\alpha} (1- n_{i\alpha})} }+ \frac{ \Delta^*_{i, \overline{\alpha}} \, d_i}{\sqrt{n_{i \overline{\alpha}} (1- n_{i \overline{\alpha}})}}
	+ \left( \Lambda_{i, \alpha} + \mu_{i, \alpha} \right) \,p_{i \alpha} 
	+ T \left( \ln\frac{|p_{i \alpha}|^2}{\Pi_{i, \alpha}} + 1\right) p_{i\alpha},
\end{eqnarray}
\begin{eqnarray}
	\label{eq:dFdd}
	\frac{\partial \mathcal{F}}{\partial d^*_i} &=& \sum_{\beta = \uparrow, \downarrow} \left[ \Delta_{i, \beta} \frac{\partial \mathcal{R}_{i, \beta}}{\partial d^*_i} 
	+  \left(\Delta_{i, \beta} \frac{\partial \mathcal{R}_{i, \beta}}{\partial n_{i, \beta}} + \Delta^*_{i, \beta} \frac{\partial \mathcal{R}^*_{i, \beta}}{\partial n_{i, \beta}} \right) 
	\frac{\partial n_{i, \beta}}{\partial d^*_i}
	+   \mu_{i,\beta} \, d_i \right]
	+ U d_i + T \left( \ln\frac{|d_i|^2}{\Pi_{i, d}} + 1\right) d_i \nonumber \\
	& & - T \sum_{\beta = \uparrow, \downarrow} \left( |e_i|^2 \frac{\partial \ln \Pi_{i, e}}{\partial n_{i,\beta}} 
	+  |p_{i\uparrow}|^2 \frac{\partial \ln \Pi_{i, \uparrow}}{\partial n_{i,\beta}} 
	+  |p_{i\downarrow}|^2 \frac{\partial \ln \Pi_{i, \downarrow}}{\partial n_{i,\beta}} 
	+  |d_i|^2 \frac{\partial \ln \Pi_{i, d}}{\partial n_{i,\beta}}  \right) \frac{\partial n_{i,\beta}}{\partial d^*_{i}} \nonumber \\
	&=& \sum_{\beta = \uparrow, \downarrow} \left[ \frac{\Delta_{i, \beta} \, p_{i, \overline{\beta}}}{\sqrt{n_{i\beta}(1-n_{i\beta})}} + \left(\Lambda_{i,\beta} + \mu_{i,\beta} \right) \, d_i \right]
	+ U d_i + T \left( \ln\frac{|d_i|^2}{\Pi_{i, d}} + 1\right) d_i
\end{eqnarray}
where $\Delta_{i,\alpha}$ is given in Eq.~(\ref{eq:Delta}), and $\Lambda_{i,\beta}$ is an effective chemical potential   
\begin{eqnarray}
	\Lambda_{i, \alpha} =  \frac{ \left(n_{i \alpha} - \frac{1}{2} \right) ( \Delta_{i, \alpha} \mathcal{R}_{i, \alpha} +  \Delta^*_{i, \alpha} \mathcal{R}^*_{i, \alpha}) }{n_{i\alpha} (1- n_{i\alpha})}
	+ T \Bigl[  |e_i|^2 \, (1 - n_{i, \overline{\alpha}}) - |p_{i, \alpha}|^2 (1-n_{i,\alpha}) + |p_{i, \overline{\alpha}}|^2 n_{i, \overline{\alpha}} - |d_i|^2 n_{i, \overline{\alpha}} \Bigr]. 
\end{eqnarray}
The set of coupled equations~(\ref{eq:dFde})--(\ref{eq:dFdd}) is highly nonlinear and has to be solved with hard constraints $|\Phi_i|^2 = 1$. A more convenient and robust approach, as outlined in Ref.~\cite{lanata12}, is to reformulate the minimization to a set of nonlinear eigenvalue problem $\mathcal{H}^{\rm SB}_i[\Phi_i] \, \Phi_i = \epsilon_i \Phi_i$, where the effective slave-boson Hamiltonian $\mathcal{H}_i^{\rm SB}$ is 
\begin{eqnarray}
	\mathcal{H}^{\rm SB}_i = \sum_\alpha \frac{ \left(\Delta_{i, \alpha} \mathcal{M}_{\alpha} + \Delta^*_{i, \alpha} \mathcal{M}^{\dagger}_{\alpha} \right) }{\sqrt{n_{i\alpha} (1- n_{i\alpha})}} + U \mathcal{U} 
	+ \sum_{\alpha} \left(\Lambda_{i,\alpha} + \mu_{i, \alpha} \right) \, \mathcal{N}_{\alpha} + T \Theta_{i}.
\end{eqnarray}
This is the same as Eq.~(\ref{eq:H_SB}) in the main text. The various matrices in the above expression are
\begin{eqnarray}
	\mathcal{M}_{\uparrow} = \left[ \begin{array}{cccc}
	0 & 0 & 0 & 0 \\
	1 & 0 & 0 & 0 \\
	0 & 0 & 0 & 0 \\
	0 & 0 & 1 & 0 
	\end{array} \right],  \quad
	\mathcal{M}_{\downarrow} = \left[ \begin{array}{cccc}
	0 & 0 & 0 & 0 \\
	0 & 0 & 0 & 0 \\
	1 & 0 & 0 & 0 \\
	0 & 1 & 0 & 0 
	\end{array} \right],  \quad
	\mathcal{U} = \left[ \begin{array}{cccc}
	0 & 0 & 0 & 0 \\
	0 & 0 & 0 & 0 \\
	0 & 0 & 0 & 0 \\
	0 & 0 & 0 & 1 
	\end{array} \right], \nonumber
\end{eqnarray}
\begin{eqnarray}
	\mathcal{N}_{\downarrow} = \left[ \begin{array}{cccc}
	0 & 0 & 0 & 0 \\
	0 & 1 & 0 & 0 \\
	0 & 0 & 0 & 0 \\
	0 & 0 & 0 & 1 
	\end{array} \right],  \quad
	\mathcal{N}_{\downarrow} = \left[ \begin{array}{cccc}
	0 & 0 & 0 & 0 \\
	0 & 0 & 0 & 0 \\
	0 & 0 & 1 & 0 \\
	0 & 0 & 0 & 1 
	\end{array} \right], \quad
	\Theta_i = \mathbb{I}_4 +  \left[ \begin{array}{cccc}
	\ln(|e_i|^2/\Pi_{i, e}) & 0 & 0 & 0 \\
	0 & \ln(|p_{i\uparrow}|^2/\Pi_{i, \uparrow}) & 0 & 0 \\
	0 & 0 & \ln(|p_{i\downarrow}|^2/\Pi_{i, \downarrow}) & 0 \\
	0 & 0 & 0 & \ln(|d_i|^2/\Pi_{i, d}) 
	\end{array} \right]. \qquad
\end{eqnarray}
The eigen-solutions of the renormalized tight-binding Hamiltonian $\mathcal{H}^{\rm qp}$ and the embedding Hamiltonians $\mathcal{H}^{\rm SB}_i$ have to be solved self-consistently. This can be achieved through the iteration procedure. Explicitly, let the solution after $k$-iteration be $\rho^{(k)}_{i\alpha, j\beta}$, $\Phi^{(k)}_i$, and $\mu^{(k)}_{i\alpha}$. The $(k+1)$-th iteration consists of the following steps: (i) New renormalization factors $\mathcal{R}_{i, \alpha}^{(k)}$ are computed from the slave-boson amplitudes $\Phi^{(k)}_i$ using Eq.~(\ref{eq:R_2}). A new tight-binding Hamiltonian $\mathcal{H}^{(k+1)}_{\rm qp} = \mathcal{H}^{\rm qp}[\mathcal{R}^{(k)}, \mu^{(k)}]$ in turn is built from these renormalization factors. (ii) KPM method is used to solve the quasi-particle density matrix $\rho^{(k+1)}_{i\alpha, j\beta}$ from the Hamiltonian $\mathcal{H}^{(k+1)}_{\rm qp}$. (iii) A new set of kinetic factors $\Delta_{i, \alpha}^{(k+1)}$ is calculated from $\rho^{(k+1)}_{i\alpha,j\beta}$ using Eq.~(\ref{eq:Delta}). The new embedding Hamiltonians are then constructed according to $\mathcal{H}^{(k+1)}_{i, \rm SB} = \mathcal{H}^{\rm SB}_i[\Phi^{(k)}, \Delta^{(k+1)}, \mu^{(k)}]$. (iv) Exact diagonalization of $\mathcal{H}^{(k+1)}_{i, \rm SB}$ then gives a new set of slave-boson amplitudes~$\Phi^{(k+1)}_i$. Finally, new values of the Lagrangian multipliers $\mu^{(k+1)}_{i,\alpha}$ are determined in order to satisfy the Gutzwiller constraints $n_{i\alpha} = \rho_{i\alpha, i\alpha}$. These steps are repeated until convergence is reached.

\section{Paramagnetic Gutzwiller solutions}
\label{sec:paramagnetic}

For non-magnetic solutions, we assume $p_{i, \uparrow} = p_{i, \downarrow}$, and $\rho_{i\alpha, j\beta} = \rho_{ij} \delta_{\alpha\beta}$. Since all derived quantities are now spin-independent, we will drop the spin indices. Futhermore, one can introduce a 3-vector for the slave-boson wavefunction $\Phi_i = (\phi_{i, 0}, \phi_{i, 1}, \phi_{i, 2}) = (e_i, \sqrt{2} p_i, d_i)$, where $p_{i} \equiv p_{i, \uparrow} = p_{i, \downarrow}$. The new wavefunction thus satisfies the same normalization condition $|\Phi_i|^2 = 1$. Using the appropriate gauge, both the renormalization factors $\mathcal{R}_i$ and the kinetic $\Delta_i$ can be made real. The free-energy now becomes
\begin{eqnarray}
	\mathcal{F}[\rho^{\rm qp}, \Phi] = 2 \sum_{\langle ij \rangle}\mathcal{R}_i \mathcal{R}_j \, t_{ij} (\rho_{ij} + \rho_{ji}) + U \sum_i D_i 
	+ 2 \sum_i \mu_i (n_i - \rho^{\;}_{ii}) 
	+ 2T {\rm Tr}(\rho^{\rm qp} \ln \rho^{\rm qp}) + T \sum_{i}\sum_{m=0}^2 \phi_{i, m}^2 \ln\frac{\phi_{i,m}^2}{\Pi_{i,m}}, \quad
\end{eqnarray}
where we have defined $\Pi_{i, 0} = \Pi_{i, e}$, $\Pi_{i, 1} = 2 \Pi_{i, \alpha}$ ($\alpha = \uparrow, \downarrow$), and $\Pi_{i, 2} = \Pi_{i, d}$. In terms of the $\phi$ variables, other relevant quantities are  \begin{eqnarray}
	n_i = \frac{1}{2} \phi_{i, 1}^2 + \phi_{i, 2}^2, \qquad D_i = \phi_{i, 2}^2, \qquad \mathcal{R}_i = \frac{\phi_{i, 0} \phi_{i, 1} + \phi_{i, 1} \phi_{i, 2}}{\sqrt{2 n_i (1- n_i)}}.
\end{eqnarray}
By expressing the $\phi_{i,m}$ in terms of $n_i$ and $D_i$ using the first two equations, one arrives at the Gutzwiller formula for the renormalization factor.
The optimization of the quasi-particle density again amounts to solving the renormalized spinless Hamiltonian
\begin{eqnarray}
	\mathcal{H}^{\rm qp} = \sum_{\langle ij \rangle} \mathcal{R}_i \mathcal{R}_j \, t_{ij} \left( c^\dagger_i c^{\;}_j + c^\dagger_j c^{\;}_i \right) - \sum_i \mu_i c^\dagger_i c^{\;}_i.
\end{eqnarray}
And similarly, the minimization with respect to the $\phi$ variables subject to condition $|\Phi_i|^2 = 1$ can be recast into the eigen-solution of the following $3\times 3$ Hamiltonian
\begin{eqnarray}
	\mathcal{H}^{\rm SB}_i = \frac{\sqrt{2}\Delta_i \mathcal{M}}{\sqrt{n_i (1-n_i)}} + U \mathcal{U} + (\Lambda_i + \mu_i) \mathcal{N} + T\Theta_i,
\end{eqnarray}
where the various matrices now become
\begin{eqnarray}
	\mathcal{M} = \left[ \begin{array}{ccc}
	0 & 1 & 0 \\
	1 & 0 & 1 \\
	0 & 1 & 0 
	\end{array} \right], 
	\quad
	\mathcal{U} = \left[ \begin{array}{ccc}
	0 & 0 & 0 \\
	0 & 0 & 0 \\
	0 & 0 & 1 
	\end{array} \right], 
	\quad
	\mathcal{N} = \left[ \begin{array}{ccc}
	0 & 0 & 0 \\
	0 & 1 & 0 \\
	0 & 0 & 2 
	\end{array} \right], 
	\quad
	\Theta_i = \mathbb{I}_3 + \left[ \begin{array}{ccc}
	\ln(\phi_{i,0}^2/\Pi_{i, 0}) & 0 & 0 \\
	0 & \ln(\phi_{i,1}^2/\Pi_{i, 1}) & 0 \\
	0 & 0 & \ln(\phi_{i,2}^2/\Pi_{i, 2}) 
	\end{array} \right], \quad
\end{eqnarray}

\section{Effective spin model}
\label{sec:spin-model}

In this Appendix, we discuss the effective spin-1/2 Hamiltonian of the Gutzwiller method based on the assumption of particle-hole symmetry and negligible charge fluctuations. This effective spin model can also be obtained using the slave-spin method for the Hubbard model. Here we consider the paramagnetic solutions with $n_{i\uparrow} = n_{i\downarrow} = 1/2$. The slave-boson amplitudes can now be represented using a pseudo-spinor $\Phi_i = (\phi_{i,1}, \phi_{i,2})$, where $\phi_{i, 1} = \sqrt{2} p_{i, \uparrow} = \sqrt{2} p_{i, \downarrow}$ and $\phi_{i, 2} = \sqrt{2} e_i = \sqrt{2} d_i$. Next we introduce spin-1/2 operator $\hat{\mathbf S} = \frac{1}{2} \bm\tau$, where $\bm\tau = [\tau_x, \tau_y, \tau_z]$ is the vector of Pauli matrices. The renormalization factors and the double-occupancy can both be written as the expectation values
\begin{eqnarray}
	\mathcal{R}_i = 2 \phi_{i,0} \phi_{i,1} = \langle \Phi_i | \tau_x | \Phi_i \rangle \equiv 2  S^x_i, \qquad
	D_i = d_i^2 = \frac{1}{2} \phi_{i,2}^2 = \frac{1}{4} \left[ 1 - \langle \Phi_i | \tau_z | \Phi_i \rangle \right] \equiv \frac{1}{4} ( 1 - 2 S^z_i ).
\end{eqnarray}
Here we have introduced classical spin variables $\mathbf S_i \equiv \frac{1}{2} \langle \Phi_i | \bm\tau | \Phi_i \rangle$. The free-energy can now be expressed as $\mathcal{F} = \mathcal{H}_{\rm eff} - T \mathcal{S}$, where the effective spin Hamiltonian and the entropy term are
\begin{eqnarray}
	\mathcal{H}_{\rm eff} = \sum_{\langle ij \rangle} \mathcal{J}_{ij} \,S^x_i S^x_j - H \sum_i S^z_i,
\end{eqnarray}
\begin{eqnarray}
	\mathcal{S} = - \sum_i \left[ \left(\frac{1}{2}+S^z_i \right) \ln\left(\frac{1}{2}+S^z_i \right) + \left(\frac{1}{2}-S^z_i \right) \ln\left(\frac{1}{2}-S^z_i \right) \right].
\end{eqnarray}
The zero-temperature part here is essentially the quantum Ising model with the ``transverse" field $H = U/2$. The effective exchange constant is
\begin{eqnarray}
	\mathcal{J}_{ij} = 8 t_{ij} (\rho_{ij} + \rho_{ji}).
\end{eqnarray}
The electrons are described by a renormalized tight-binding Hamiltonian $\mathcal{H}^{\rm qp} = \sum_{\langle ij \rangle} t^{\rm eff}_{ij} (c^\dagger_i c^{\;}_j + \mbox{h.c.})$, where the effective hopping depends on the spin configuration, i.e. $t^{\rm eff}_{ij} = 4 S^x_i\, S^x_j \, t_{ij}$.
In the spin picture, the order parameter of the Mott transition is the $z$-component of the average spin; the first-order Mott transition is marked by a jump in $\langle S^z \rangle$ when crossing the transition temperature from the correlated metal to the Mott phase. Importantly, the nearest-neighbor exchange constant $\mathcal{J}_{ij}$ depends implicitly on spins at further neighbors through the nonlocal electronic degrees of freedom $\rho_{ij}$.

\twocolumngrid

\newpage


\begin{thebibliography}{99}
\bibitem{note-dUdT} The slope of $dU_c/dT$ is shown to be positive for square lattice when inter-site correlations are included in the cluster DMFT calculations. However, as shown in Ref.~\cite{liebsch09}, cluster DMFT calculation on the frustrated triangular lattice finds a phase boundary with negative~$dU_c/dT$.
\end{thebibliography}
\end{document}